\documentclass[aps,twocolumn,superscriptaddress,altaffilletter,lengthcheck,showpacs, tightenlines,showkeys]{revtex4}
\usepackage{graphicx}
\usepackage{color}
\usepackage{amstext}
 \usepackage[dvipdf]{epsfig}
  \usepackage{color}
  \usepackage{amssymb}
\def\k{\textsf{k} }
\def\q{\textsf{q} }
\def\no{\nonumber}
\def\lb{\label}
\def\be{\begin{equation}}
\def\ee#1{\label{#1}\end{equation}}

\newcommand{\ben}{\begin{eqnarray}}
\newcommand{\een}{\end{eqnarray}}

\begin{document}
\title{Jeans Instability in a Universe with Dissipation}

\author{Gilberto M. Kremer}\email{kremer@fisica.ufpr.br}
\affiliation{Departamento de F\'isica, Universidade Federal do Paran\'a, Caixa Postal 19044, 81531-990 Curitiba, Brazil}
\author{Mart\'{\i}n G. Richarte}\email{martin@df.uba.ar}
\affiliation{Departamento de F\'isica, Universidade Federal do Paran\'a, Caixa Postal 19044, 81531-990 Curitiba, Brazil}
\affiliation{Departamento de F\'isica, Facultad de Ciencias Exactas y Naturales,
Universidad de Buenos Aires, Ciudad Universitaria 1428, Pabell\'on I,  Buenos Aires, Argentina}
\author{Felipe Teston}\email{felipe@fisica.ufpr.br}
\affiliation{Departamento de F\'isica, Universidade Federal do Paran\'a, Caixa Postal 19044, 81531-990 Curitiba, Brazil}

\bibliographystyle{plain}

\begin{abstract}
The problem of Jeans gravitational instability is investigated for static and  expanding universes within the context of the five and thirteen field theories which account for  viscous and thermal effects. For the five-field theory a general dispersion relation has been derived with the help of relevant linearized perturbation equations, showing that   the shear viscosity parameter alters the propagating  modes for large and small wavelengths. The behavior of  density and temperature contrasts are analyzed for the hard-sphere model in detail. In the  small wavelengths regime,  increasing the amount of shear viscosity  into the system  forces the harmonic perturbations to damp faster, however,  in the opposite limit larger values of shear viscosity lead to smaller values of density and temperature contrasts. We also consider the hyperbolic case associated with the thirteen-field theory  which involves two related parameters, namely  the shear viscosity and the collision frequency, the last one is due to the production terms  which appear in the Grad method.  The dispersion relation  becomes a polynomial in the frequency with two orders higher in relation to the five-field  theory, indicating that the effects associated  with  the shear viscosity and heat flux  are non-trivial.   The profile of Jeans mass in terms of  the temperature and number density is explored by contrasting with several data of molecular clouds. Regarding the dynamical evolution  of  the density, temperature, stress  and heat flux contrasts for a universe dominated by pressureless matter, we obtain  also  damped harmonic waves for small wavelengths. In the case of large wavelengths, the density and temperature contrasts grow with time  (due to the Jeans mechanism) while the stress and heat flux contrasts heavily decay with time. 
\end{abstract}

\vskip 1cm

\date{\today}
\maketitle

 \section{Introduction}
Current view of the universe seems to indicate that it is vastly uniform and homogeneous at large-scale for redshifts larger than $z \simeq 10^3$ \cite{DM1}, \cite{DM2}.
However,  the existence of inhomogeneities such as galaxies and clusters of galaxies requires  a physical mechanism to
account them for. One way to address the study of structure formation is by considering the process associated with the
aggregation of matter in a form of cosmic substratum (or fluid), which permeate the whole universe. From some initial seeds,
possibly generated during a primordial inflationary era, one must explore how gravity forces  to  grow into lumpy structures
on small scale that one observes today, namely a clustered  distribution of galaxies or super-cluster of galaxies at low redshifts ($z \ll 1$) \cite{DM2}.

First attempts to examine the existence of instability for a fluid  within the context of collapsing astrophysical bumps
were initiated by  Sir Jeans in 1912 \cite{Jeans}, \cite{Jeans2}. He started his analysis by assuming a static universe and found that the dispersion
relation associated with the perturbed contrast density could admit, apart from harmonic perturbations, two kinds of propagating modes: a growing  mode  and a decaying one. To be more precise, he noticed the existence of a physical cutoff, called Jeans' wavelength, such that  perturbations with
wavelength  shorter  than Jeans scale  will not grow whereas in the opposite case perturbations  may grow exponentially in time.
A useful manner to understand the gravitational instability associated with the Jeans mechanism  is by using the following reasoning: consider a given mass $M$ enclosed in a spherical volume of radius $\bm{\lambda}$ in which there exists a mass density inhomogeneity. This inhomogeneity will grow if the gravity force $F_G$ per unit mass is greater than the opposed pressure force $F_P$ per unit of mass, i.e.,
\[
F_G= \frac{GM}{\lambda^2}\propto \frac{G\rho\lambda^3}{\lambda^2}=G\rho\lambda>F_P\propto \frac{p\lambda^2}{\rho\lambda^3}\propto \frac{v_s^2}\lambda,
\]
where one used  $v_s^2\propto (p/\rho)$. $G$ stands for the Newton gravitational constant, $\rho$ represents the  density,  and $\lambda$ is a wavelength. From dimensional analysis, one  shows that the Jeans wavelength  can be written in terms of the Jeans wave-number as $\lambda_J=2\pi/\k_J={2\pi v_s}/{\sqrt{4\pi G\rho}}$ being $\lambda=2\pi/\k$; then instability arises if the condition $\k<\k_J$ holds. Equivalently, the Jeans instability can be stated as follows: the timescale associated with the pressure exerted over a region with matter must be bigger than the timescale needed to start the gravitational collapse of the matter due to its own weight, namely $t_{\rm pressure}=(\lambda/v_{s})>t_{\rm gravity}=(G\rho)^{-1/2}$.

The situation of an expanding universe is considerably different provided the  expansion tries to counter-balance the effect of gravity.
As a positive density bump is trying to collapse due to its self-gravity the expansion of the universe is trying to pulling it apart \cite{Bono}.  In fact, there are two  elements which can help to prevent the formation of structure, which are the usual  pressure effects  and the fast background expansion.
Therefore,  the growth rate of matter in an expanding universe tends to be  slower than in one with no expansion, evolving as power
law in time in the latter case. The Jeans treatment for exploring the evolution of density contrast is just limited to the linear regime,
so as the amplitude of density fluctuations approach to the non-linear stage all the propagating modes do not evolve independently.
Nevertheless, Jeans stability analysis seems to be the first method to understand the inclusion of secondary effects in the agglomeration of matter
for the non-expanding case and expanding scenario  within the framework of first-order Newtonian dynamics. Some authors devoted several efforts to explore the Jeans instability mechanism within the context of Newtonian cosmology where the
universe is dominated by a pressureless matter. For instance, the effect of including a non vanishing bulk viscosity   along with
the  analysis of the top-down fragmentation process  were taken into account in Ref. \cite{monta}. In doing so, they compared one collapsing agglomerate with $M\gg M_{J}$ and an internal non-collapsing sub-structure with $M < M_{J}$, $M_{J}$ being the Jeans mass \cite{monta2}.

The case of Jeans instability criterion for
a compressible fluid including viscosity and heat conduction was analyzed many
years ago \cite{coro}. At high redshifts associated with the recombination epoch ($z \simeq 1000$) the role played by the shear viscosity would be negligible  while the bulk viscosity would be important \cite{coro}. In addition,  the gravitational instability for a fluid  which supports viscoelastic stresses was addressed by Janaki \emph{et al.}, showing that quadratic corrections appear in the dispersion relation due to the presence of  both shear viscosity and bulk viscosity \cite{jana}.

On the other hand, the analysis of Jeans instability from the point of view of kinetic theory was also examined in the literature. In such treatment, Trigger \emph{et al.} obtained the dispersion relation for the Boltzmann-Vlasov operator coupled to Poisson  equation.  The authors  showed that a non-vanishing collisional term led to substantial modifications to the unstable propagating modes \cite{trigger}. The  gravitational instability for a collisionless self-gravitating system composed of one or two components described by the coupled collisionless Boltzmann and Poisson equation  were analyzed   as a first step to understand the  cooperative effects  of baryons and dark matter in  process of structure formation \cite{kremer1}, \cite{kremer2}. For instance,   a system composed of  baryons and dark matter lead to a total Jeans
mass  which is smaller than the one associated with a single component, pointing out that less amount of mass is needed to ignite the collapsing process. One could expect that for  bumps with masses greater  the Jeans mass initiate the collapsing process but an overdense region in  an expanding universe eventually recollapses and  virialises. In the case of a single component with an expanding background it turned out that the ``swindle'' proposal  may be avoided while the Jeans instability is expected to arise in the limit of large wavelengths  \cite{kremer1}.  Leaving General Relativity aside, some authors devoted  several efforts to explore the Jeans instability within the framework of $f(R)$  gravity or modified gravity theories \cite{raila} , \cite{dolgo}, \cite{capo1}, \cite{capo2}, \cite{capo3}.
In particular, Capozziello \emph{et al.} studied  the collapse of  self-gravitating system composed of  neutral particles which is characterized by a coupled of collisionless Boltzmann and Poisson equations in the case of $F(R)$-gravity. Perturbing a background solution and looking at linearized regime (or in the case of weak field approximation) of the aforesaid set of  equations, they obtained  a modified dispersion relation along with a new kind of an unstable mode. Such propagating mode turned to grow faster than the standard one obtained in GR \cite{capo2},  \cite{capo3}.

The aim of this work is to explore the Jeans instability for a static and expanding backgrounds from the phenomenological point of view of a five-field and a thirteen-field theory, so it can be considered as a complementary analysis to the full kinetic theory approach mentioned above. First, we are going to  consider  a proper fluid-description in terms of a five-field model  which is characterized by several fields such as  mass density, velocity field,  temperature, gravitational potential along with an equation of state to close the coupled Poisson-balance system of equations. In the thirteen-field theory, we are going to deal with  a hyperbolic system of equations which include mass density, velocity, pressure tensor and heat flux vector. For an ideal gas these balance equations can be obtained from the Grad method of kinetic theory \cite{GK} or from the so-called extended thermodynamic theory (see \cite{Mu}). The main idea of such approach is
to analyze the evolution of plane waves and obtained the modified dispersion relation for static universe. Then, we are going to solve numerically the dynamical set of equations for inspecting the evolution of contrast density, contrast temperature, contrast heat flux and contrast shear tensor with time  when the expanding universe is dominated by dust-like matter. In doing so,   we will explore the impact and effect introduced by the shear viscosity parameter in the evolution of contrast quantities such as density, temperature,  heat flux and  shear's stress. In the phenomenological side of our model, we will examine two different kinds of constraints.  We will explore for  static configurations how the Jeans mass varies with the temperature and number density and compare such outcomes with several data of molecular clouds. In addition, we will study the dynamical Jeans mass in terms of the redshift.

The outline of this paper is as follows. Sec. IIA is dedicated to analyze the Jeans instability  for static universe along with  the computation of the dispersion relation (and growing mode) within the five-field theory while in Sec. IIB several constraints on the static Jeans mass are obtained in the case of stellar molecular clouds.  Sec. IIC is devoted to evaluate the evolution of contrast quantities (namely, mass density, temperature)  in terms of the cosmic time.
In Sec. IID the dynamical Jeans mass is calculated to estimate  the clustering properties of the model.
In Sec.IIIA, the  thirteen-field theory is presented, the dispersion relation in the presence of shear viscosity is calculated. In Sec.IIIB, the dynamical evolution of thirteen fields  coupled to the Poisson equation is explored along with the physical behavior of such quantities with the shear viscosity. In Sec. IV, the conclusions are stated.

\section{Jeans instability from a five-field theory}

The aim of ordinary thermodynamics  of fluids is the determination of five fields of mass density $\rho$, velocity $v_i$ and temperature $T$ in space-time $(t,x_i)$. The  equations for the determination of these fields are based on the balance equations of mass, momentum and energy which in the presence of a gravitational potential $\phi$ read (see e.g. \cite{Mu,GK})

\ben\no\frac{\partial \rho}{\partial t}+\frac{\partial \rho v_i}{\partial x_i}=0,
\quad
\frac{\partial v_i}{\partial t}+v_j\frac{\partial  v_i}{\partial x_j}+\frac1\rho\frac{\partial p_{ij}}{\partial x_j}+\frac{\partial \phi}{\partial x_i}=0,\\\lb{1a}
\frac{\partial \varepsilon}{\partial t}+v_i\frac{\partial \varepsilon}{\partial x_i}+\frac1\rho\frac{\partial q_i}{\partial x_i}+\frac{p_{ij}}\rho\frac{\partial v_i}{\partial x_j}=0.
\een
In the above equations $p_{ij}$ denotes the pressure tensor, $\varepsilon$ stands for the specific internal energy and $q_i$ is the heat flux vector.

The gravitational potential is connected with the mass density through Poisson's equation, namely
\ben\lb{1b}
\nabla^2\phi=4\pi G\rho,
\een
where $G$ is the gravitational constant. In order to close the system of equations (\ref{1a}) and (\ref{1b}) constitutive equations for  $p_{ij}, \varepsilon, q_i$ in terms of the basic fields $\rho, v_i, T$ must be specified. Here we are interested in studying an ideal gas where  velocity and temperature gradients are present. According to the thermodynamic theory the constitutive equations for the pressure tensor and heat flux vector are governed by the so-called Navier-Stokes and Fourier laws and their expressions read

\ben p_{ij}=p\delta_{ij}-\mu_{v}\left(\frac{\partial v_i}{\partial x_j}+\frac{\partial v_j}{\partial x_i}-\frac23\frac{\partial v_r}{\partial x_r}\delta_{ij}\right),\\\lb{2a}
 q_i=-{\lambda}_{c}\frac{\partial T}{\partial x_i},
\een
respectively. Above $p=\rho k T/m$ is the pressure with $k$ denoting the Boltzmann constant and $m$ the mass of a gas particle, $\mu_{v}$ is the shear viscosity and $\lambda_{c}$ the thermal conductivity coefficients, respectively. According to the kinetic theory for an ideal non-relativistic gas \cite{GK} the bulk viscosity vanishes and the coefficients of shear viscosity and thermal conductivity are related by  the following expression:
\ben\lb{cvs}
\frac{\lambda_{c}}{\mu_{v}}=\frac{15k}{4m}.
\een
Furthermore, for an ideal gas the specific internal energy is given by $\varepsilon=3kT/2m$.

In the next subsections, we shall investigate the Jeans instability for a static and expanding Universe by taking into account the above equations.

\subsection{Static Universe}

Here we search for  plane wave solutions of (\ref{1a}) and (\ref{1b}) supplemented by the constitutive equations (\ref{2a}). At equilibrium the density, temperature and gravitational potential have constant values, namely $\rho_0, T_0, \phi_0$ while the velocity vanishes. The equilibrium solution is subjected to perturbations characterized by a Fourier expansion with frequency $\omega$, wavenumber vector $\k$  and small amplitudes $\{\delta\rho, \delta v,\delta T, \delta \phi\}$ in relation to unperturbed variables:
\ben\lb{3a}
\rho=\rho_0+\delta\rho e^{i\left(\k x-\omega t\right)},\quad
v_x=\delta v_xe^{i\left(\k x-\omega t\right)},
\\\lb{3b}
T=T_0+\delta Te^{i\left(\k x-\omega t\right)},\quad
\phi=\phi_0+\delta\phi e^{i\left(\k x-\omega t\right)}.
\een
For simplicity, we have assumed that $\bar{k}=\k \hat{x}$. Further, it is import to remark  the fact that the equilibrium values for the fields satisfy the  equations (\ref{1a}) but not Poisson equation (\ref{1b}). This inconsistency is removed by considering
 Jeans ``swindle" approximation which states that the Poisson equation is valid only for the perturbations.

Replacing the Fourier expansion modes (\ref{3a}) and (\ref{3b}) together with the constitutive equations (\ref{2a}) into the balance (\ref{1a}) and Poisson (\ref{1b}) equations, one  obtains  a system of algebraic equations for the amplitudes. This system of algebraic equations has a solution if the determinant associated with  the  coefficients of the perturbations vanishes and consequently  one gets the following dispersion relation
\ben\no\omega\left(\omega +i\frac{4\mu_v}{3\rho_0} \k^2\right)\left(\omega+i\frac{5\mu_v}{2\rho_0}\k^2\right)-\frac{2k}{3m}T_0\omega \k^2
\\\label{4}
-\left(\omega+i\frac{5\mu_v}{2\rho_0}\k^2\right)\left(\frac{k}mT_0\k^2-4\pi G\rho_0\right)=0,
\een
where $\mu_{v}$ stands for the viscosity coefficient which is a function of the temperature \cite{chap}. Note that one has to use the relationship between the coefficients of thermal conductivity and shear viscosity (\ref{cvs}) in order to  recast (\ref{4}) in its current form.

Now we shall  analyze  the dispersion relation (\ref{4}) and for that end we introduce
 the adiabatic speed of sound $v_s$, the Jeans wavenumber $\k_J$ and the  dimensionless parameters associated with the shear viscosity coefficient  $\mu_*$, frequency $\omega_*$ and wavenumber $\k_*$:
\ben
v_s=\sqrt{\frac{5k}{3m}T_0},~~~~~~
\k_J=\frac{\sqrt{4\pi G\rho_0}}{v_s},  
\\\lb{5}
\mu_*=\frac{\mu_v \k_J}{\rho_0v_s},~ \k_*=\frac{\k}{\k_J}=\frac{\lambda_J}{\lambda},~
\omega_*=\frac\omega{v_s\k_J},
\een
where $\lambda_J=2\pi/\k_J$ is the Jeans wavelength.

In terms of the dimensionless quantities (\ref{5}) equation (\ref{4}) can be rewritten as
\ben\no\omega_*^3+i\frac{23}{6}\mu_*\k_*^2\omega_*^2+\left(1-\k_*^2-\frac{10}{3} \mu_*^2 \k_*^4\right)\omega_* \\\lb{6}
+i\frac{5}2\mu_*\k_*^2\left(1-\frac35\k_*^2\right)=0.
\een

At this point, one must look for  an important consistency check which corresponds to the case of  a non-viscous and non-thermal conducting gas or equivalently $\mu_*=0$.  In that limit  (\ref{6}) leads to
\ben\lb{7a}
\omega_*=\pm\sqrt{\frac{\lambda_J^2}{\lambda^2}-1},
\een
which is the well-known Jean's solution. For small wavelengths $\lambda_J>\lambda$, one finds that $\omega_*$ remains a real quantity. Due to the factor $\exp(-i\omega t)$ this perturbations will propagate as harmonic waves in time. For big wavelengths $\lambda_J<\lambda$,  $\omega_*$ becomes a pure imaginary quantity then the perturbations will grow or decay in time, depending on the $\pm$ sign. Further,  the one which grows in time is connected with Jeans instability. Now it is essential to compare the dispersion relations in the case $\mu^{*}=0$ and $\mu_*\neq 0$. We check that  dimensionless dispersion relation in the presence of dissipative effects is smaller than the dispersion relation with $\mu^{*}=0$ [cf. Fig. (\ref{fig1})]. This is physically equivalent to state that the time needed to collapse a mass configuration is larger when the dissipative effects are included.
\begin{figure}[h!]
\includegraphics[width=16cm]{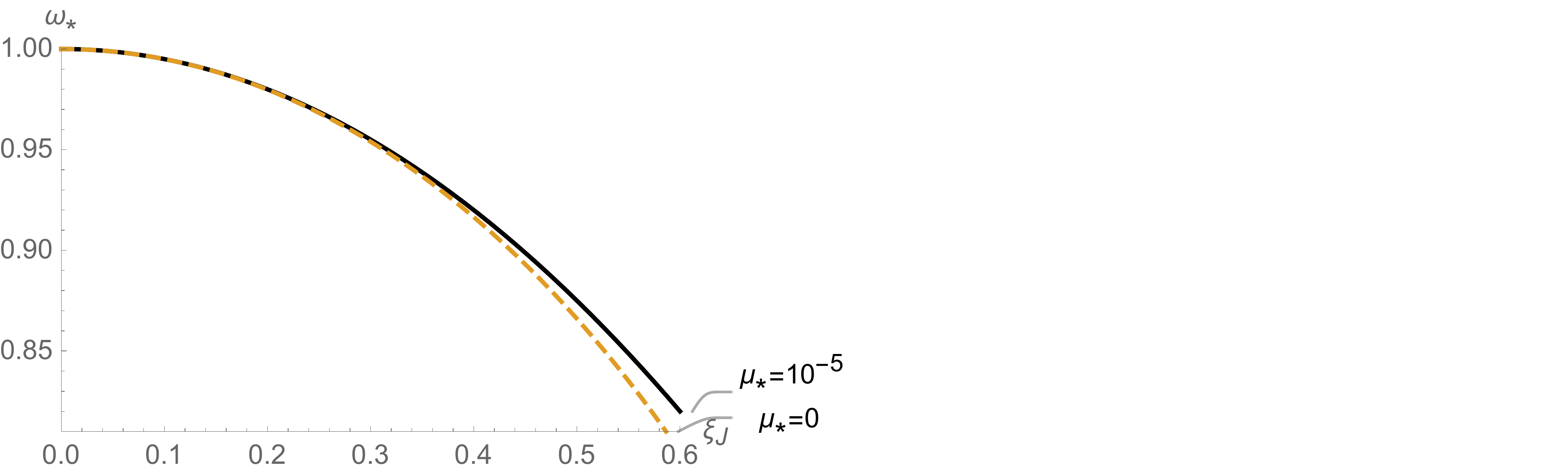}
\caption{Behavior of the dimensionless dispersion relations in terms of $\xi_{J}=\lambda_{J}/\lambda$ for two different cases, namely,  $\mu^{*}=0$ and $\mu_*=10^{-5}$.}
\lb{fig1}
\end{figure}

For  large wavelengths $\lambda_J/\lambda<1$, we can obtain from (\ref{6}) three different values associated with the dimensionless frequencies that read

\ben\omega_{*1}=-i\left[\frac{5\mu_*}2\left(\frac{\lambda_J}{\lambda}\right)^2+\dots\right],\\
\omega_{*2}=-i\left[1-\frac{3-4\mu_*}{6}\left(\frac{\lambda_J}{\lambda}\right)^2-\dots\right], \\
\lb{7b}
\omega_{*3}=i\left[1-\frac{3+4\mu_*}{6}\left(\frac{\lambda_J}{\lambda}\right)^2-\dots\right].
\een
These dimensionless frequencies are pure imaginary. In particular,  $\omega_{*1}$ and $\omega_{*2}$ result in decaying perturbations in time while $\omega_{*3}$ represents a growing mode. This last refers to Jeans instability which shows a dependence on dissipative phenomena through the dimensionless shear viscosity coefficient $\mu_*$. 
For  small wavelengths $\lambda_J/\lambda>1$, we obtain two complex conjugate solutions for the dimensionless frequencies which refer to harmonic waves in time and a decaying mode.
\subsection{Stellar structure formation}
In this subsection, we are going to  explore the impact of viscosity   within the context of stellar structure formation as a useful application of the previous analysis.  Galaxies have different components  such as stars  and  the interstellar medium.  In particular, the  interstellar medium contains clouds of cold and ionized gas and dust, like giant molecular clouds, HII regions, Bok globules, etc. These clouds are usually stable, with pressure (due to a finite temperature) balancing self-gravity.  In general, the physical processes of a collapsing clouds are well described in the fluid limit, however,  it can not take into account some effects due to the particle nature of the interstellar medium  that could much be larger if an alternative gravity theory is selected \cite{capo2}.  In order to compare  the standard Jeans mass with the case of non-zero viscosity,  we must begin by obtaining the Jeans masses in both cases.  In the standard case with $\mu_{*}=0$, the Jeans m
 ass can be visualized as the minimum mass for an overdensity to collapse.  It is defined as the mass contained within a sphere of diameter $\lambda_{J}$, so it reads $M^{s}_{J}=(4\pi/3) (\lambda_{J}/2)^3 \rho_{0}$,  where the mass density is $\rho_{0}$. We assume that the main component of these clouds is hydrogen ($H_{2}$), so the density is given by $\rho_{0}=\mu m_{p}n$, $n$  being the number density, $m_p$ stands for the proton mass and $\mu=2.06$ is the mean molecular weight for  Hydrogen.   For an ideal gas, the squared thermal velocity is proportional to the temperature of the cloud, $v^{2}_{s} \propto kT/m_{H_{2}}$, so with the help of (\ref{5}) the Jeans mass can be recast as
\ben\lb{ma1}
M^{s}_{J}= \frac{\pi^4}{6(\mu m_{p})^{2}n^{1\over2}}\Big(\frac{5}{12\pi} \frac{kT}{G} \Big)^{3\over2}.
\een
Eq.  (\ref{ma1})  can be written  is solar units as $M^{s}_{J}\simeq 3.31~ T^{3\over2} n^{-\frac{1}{2}} M_{\odot}$, where $T[{\rm K}]$ is the numerical value of temperature in Kelvin units and $n [\rm {cm}^{-3}]$ the number density in centimeters.   At this point some comments are in order. When we are dealing with the structure formation at the astrophysical level there are several mechanisms which we must take into account. For instance, we must include the fragmentation/starvation process, the radiation pressure, stellar winds and so on.  Here we are studying an unstable media which collpase gravitationally and how they produce objects with a characteristic mass comparable to the Jeans mass (see \cite{CL}).
Given the fact that the great majority of stars form via the collapse of cold, gravitationally-unstable, molecular gas, and the subsequent accretion of cold gas onto the protostellar seeds that the collapse produce, it is quite natural to estimate the Jeans mass:
\ben\lb{meff}
M^{s}_{J}= 0.5\Big(\frac{n}{10^{4}{\rm cm}^{-3}}\Big)^{1\over2} \big(\frac{T}{10 {\rm K}} \big)^{3\over2}M_{\odot}
\een
Eq. (\ref{meff}) shows  the typical values of the temperature and  the number density associated with star-forming regions. The general mechanism of how a gas does fragment into numerous small stars rather than forming a single large one is a really complex process.   Nevertheless we can mention the standard fragmentation mechanism. As can be seen from (\ref{meff}), the critical mass is only a function of temperature and density. Increasing density and decreasing temperature lead to lower Jeans masses which, in turn, will lead to fragmentation into smaller cores.  For a given initial mass, a molecular cloud (or other astrophysical object) may be formed in spiral density waves and other density perturbations (e.g., caused by the expansion of a supernova remnant or superbubble). What exactly happens during the collapse depends very much on the temperature evolution of the cloud. Initially, the cooling processes (due to molecular and dust radiation) are very efficient. If the cool
 ing time scale is much shorter than  free fall time, the collapse is approximately isothermal.
As $M^{s}_{J}$ decreases, inhomogeneities with mass larger than the actual $M^{s}_{J}$ will collapse by themselves with their local free-fall time.
This fragmentation process will continue as long as the local cooling time is shorter than the local free-fall time, producing increasingly smaller
collapsing subunits. Eventually the density of subunits becomes so large that they become optically thick and the evolution becomes adiabatic.
As the density has to increase, the evolution will always reach a point when $M=M^{s}_{J}$, when a subunit reaches approximately hydrostatic equilibrium. We assume that a stellar object is born \cite{pad}. At this point, we should remark the main reason for studying molecular clouds. As is well known, the primary sites for star formation are molecular clouds. These are thought to form out of the remnants of supernova explosions, outflows and winds from stars,
and the gas reservoirs in the interstellar and the intergalactic medium.  Besides, we should also stress that from a giant molecular cloud it can form a group of stars with their mass distribution being determined by the fragmentation process. The process depends on the physical and chemical properties of the cloud (ambient pressure, magnetic field, rotation, composition, dust fraction, stellar feedback, etc.); that is, much of the fragmentation process  is still under study and is not completely understood.

Our aim is to compare  the Jeans mass of the interstellar gas with our without dissipative effects.  Then, the study of  the ionization feedback, the radiation pressure, and stellar wind in process of fragmentation/starvation  are beyond the scope of the present work. In the case of non-vanishing viscosity, we use (\ref{5}) along with the definition of Jeans mass to exhibit how such quantity scales  with the viscosity, it reads
\ben\lb{ma2}
M^{\mu}_{J}= \frac{\pi^4}{3\mu^{3}_{*}} \rho^{-2}_{0}\mu^{3}_{v}v^{-3}_{s}.
\een
To consider the viscosity effects as subleading ones we must choose $\mu_{*}$ considerably small, typically of order  $10^{-6}$ or  $10^{-7}$ otherwise we can obtain unrealistic scenarios.  As is well known the viscosity   depends on the temperature of the gas, so we have to select a model for $\mu_{v}(T)$. We choose the  toy model of Maxwell given by $\mu_{v}=3.1 \times 10^{-7} T~ [{\rm g}~{\rm cm}^{-1} {\rm s}^{-1} {\rm K}^{-1}]$, where once again we assumed that the gas is entirely composed of hydrogen (see \cite{chap}).  From the previous choice, we obtain that the Jeans mass behaves as $M^{\mu}_{J}\simeq 3.79 \times 10^{-17} T^{3\over2} n^{-2} \mu^{-3}_{*} M_{\odot}$. Before embark us in the numerical analysis of the Jeans mass in both scenarios, we begin by noting several points. Eqs. (\ref{ma1})-(\ref{ma2}) tell us that Jeans mass scales as $T^{3/2}$  so  the $M_{J}-T$ relation for the interstellar medium is the same in both cases. Nevertheless, the behavior with
 the numerical density is different in both cases  and in addition the viscous case contains the parameter $\mu_{*}$. It is illustrative  to consider different astrophysical systems to compare the Jeans mass in these two cases \cite{capo2}.
\begin{itemize}
\item \textbf{Giant molecular clouds} are composed of gas, dust and different substructures extended over a region of $10 {\rm pc}$. They exhibit a high number  density $ 10^{2}-10^{3} {\rm cm}^{-3}$  and a very low-temperature, $10 [{\rm{K}}]$. Their masses vary between   $10^{4}M_{\odot}$ and $10^{6}M_{\odot}$.
 \item \textbf{Cold neutral medium } is entirely composed of neutral hydrogen  and exhibits a low number  density $ 30{\rm cm}^{-3}$  and a very low-temperature, $10-100 [{\rm{K}}]$.
  \item \textbf{Warm neutral medium} is composed of neutral hydrogen ${\rm HI}$  with a temperature  of $10^{3}- 10^{4} [{\rm{K}}]$ and a low density $n \simeq 0.6 {\rm cm}^{-3}$.
  \item \textbf{Warm ionized medium} is a considerably large low-density photo-ionized zone with  with a temperature of  $8 \times 10^{3} [{\rm{K}}]$, bigger than the sun's temperature at its surface. It has a lower density than the warm neutral medium, namely $n \simeq 0.1 {\rm cm}^{-3}$.
  \item \textbf{HII regions} are interesting objects in the galaxy provided   they basically emit in  the radio and IR part of the visible spectrum, then they are easy to explore observationally. They have an average temperature around $10^{4} [{\rm{K}}]$ and  a typical density in the range of  $ 0.1-10^{4} {\rm cm}^{-3}$. The mass of HII regions varies between $10^{2}M_{\odot}$ and $10^{4}M_{\odot}$.
  \item \textbf{Hot intercloud region} is composed of ionized gas ${\rm HII}$ with  high temperatures,$10^{5}- 10^{6} [{\rm{K}}]$. Such region is extended over $20 {\rm pc}$ with an extremely low density of  $ 0.004 {\rm cm}^{-3}$.
   \item \textbf{Intracluster medium}  exhibits  high temperatures, $10^{7}- 10^{8} [{\rm{K}}]$,  a very low density $ 10^{-3} {\rm cm}^{-3}$ and is extended over a a few Mpc.

  \item \textbf{Fermi Bubbles} are extended over $10 {\rm kpc}$ with a temperature that varies from $10^{8} [{\rm{K}}]$  to $10^{9} [{\rm{K}}]$, exhibiting a density of $ 0.01 {\rm cm}^{-3}$.  These configurations are located above and below the galactic plane.
\end{itemize}

\begin{table}[!h]
\centering
\begin{tabular}{||c c c c c ||}
\hline
Object & {\rm T} $[{\rm K}]$ & {\rm n}  $[{\rm cm}^{-3}]$ &  $M^{\mu}_{J} [M_{\odot}] $ & $M^{s}_{J}[M_{\odot}]$\\ [0.5ex]
\hline\hline
GMC &  10& 100 & 120.15  &  10.49 \\
CNM & 80& 30 & 30207  &  433.44 \\ [1ex]
WNM & 8000& 0.6& $7.55~ 10^{10}$ & $3.06 ~ 10^{6}$ \\ [1ex]
WIM & 8000& 0.1& $2.71~ 10^{9}$ & $7.51 ~ 10^{6}$ \\ [1ex]
HII & $10^4$ & 0.1 & $3.79~ 10^{9}$ & $1.04 ~ 10^{7}$ \\ [1ex]
HIR &  $10^4$&  $4~ 10^{-3}$ & $2.37 ~10^{15}$ & $5.24 ~ 10^{10}$ \\ [1ex]
ICM & $10^{7}$& $10^{-3}$ & $1.20~10^{18}$ & $3.31 ~ 10^{12}$ \\ [1ex]
FB & $10^{8}$& $10^{-2}$ & $3.79~ 10^{17}$ & $3.31~ 10^{13}$ \\ [1ex]
\hline
\end{tabular}
\caption{Jeans mass measured in solar units for different values of temperatures and number densities.}
\label{tab1}
\end{table}
As is expected the Jeans mass for a system with a small viscosity gives always greater mass than the case without viscosity. For instance, the smallest difference corresponds to the case of  giant molecular clouds, being  $M^{\mu}_{J}/M^{s}_{J} \simeq 11.45$, while the largest disagreement appears for the intracluster medium provided $M^{\mu}_{J}/M^{s}_{J} \simeq 3.6 \times 10^{5}$.  The previous result can be easily understood by taking into account  that the temperature factor is the same in both cases, but the behavior with the number density is stronger  for the viscosity case. Further, we also have to  add the  $\mu^{-3}_{*}$ term, which accounts for a $10^{21}$ factor in the Jeans mass with viscosity [cf. Tab. (\ref{tab1})]. At this point, two comments are in order regarding  the high values associated with the Jeans' mass for HIR, ICM and FB [cf. Tab. (\ref{tab1})]. Clearly these astrophysical objects are observed in the galaxy; however, the fragmentation process that leads to such configurations  endowed with the small values of viscosity seems to be very unlikely. So, we think that the small viscosity condition  breaks down for such objects and we need to consider a scenario where the viscosity is extremely high, for instance, recent numerical analysis showed that FBs are well described in terms of high viscosity values \cite{Fermi}. In addition to that, it is possible that  we should change  the  phenomenological law adopted for $\mu_{v}(T)$ \cite{Fermi}.

In general,  we expect that the system with viscosity leads to less effective system regarding the aggregation of mass  and the posterior collapse of a gravitational cloud provided the viscosity and conductive work against the  structure form
 ation process.  To reinforce such idea, we are going to explore another kinds of molecular clouds which are characterized by  really low temperature, $T \leq 10 {\rm{K}}$ and slightly high number density, that is,  $n \leq  10^{2} {\rm cm}^{-3}$ [cf. Tab.\ref{tab2}]. In doing so, we will use some of the samples mentioned in Capoziello \textit{et al. }\cite{capo2} which were extracted from Ref. \cite{jack}.

\begin{table}[h!]
\centering
\begin{tabular}{||c c c c c ||}
\hline
Object & {\rm T} $[{\rm{K}}]$ &~~ {\rm n}  $10^{2}[{\rm cm}^{-3}]$ &  $M^{\mu}_{J} [M_{\odot}]$ &~ $M^{s}_{J}[M_{\odot}]$\\ [0.5ex]
\hline\hline
GRSMC G053.59 & 5.97& 1.489 &25.30 &3.97 \\
GRSMC G049.49 & 6.48& 1.54  &26.42 &4.41 \\
GRSMC G018.89 & 6.61 & 1.58 &25.86 & 4.48\\
GRSMC G030.49 & 7.05 & 1.66 &25.81 &4.82\\
GRSMC G035.14 & 7.11 & 1.89 &25.81 &4.82\\
GRSMC G034.24 &7.15  & 2.04 &20.16 &4.57\\
GRSMC G019.94 & 7.17 & 2.43 &17.45 &4.41\\
GRSMC G038.94 & 7.35 & 2.61 &12.35 &4.08\\
GRSMC G053.14 & 7.78 & 2.67 &11.11 &4.09\\
GRSMC G023.24 & 8.57 & 3.75 &11.56 & 4.40\\
GRSMC G019.89 & 8.64 & 3.87 &6.77  & 4.29\\
GRSMC G022.04 & 8.69 & 4.41 &6.44  & 4.28\\
GRSMC G018.89 & 8.79 & 4.46 &4.96  &4.04 \\
GRSMC G023.34& 8.87 & 4.99 &4.97  &4.09 \\
\hline
\hline
\end{tabular}
\caption{Temperature, number density and Jeans masses with and without the dissipative effects  for some molecular clouds reported in \cite{jack}. For the sake of brevity, we shortened the name of the above objects.}
\label{tab2}
\end{table}
Interestingly enough, Tab. (\ref{tab2}) tells us that the largest difference appears in the case of   $T=6.48 {\rm K}$ and $n= 1.54 \times 10^{2} {\rm cm}^{-3}$, which leads to a ratio $M^{\mu}_{J}/M^{s}_{J} \simeq 5.99$ while the smallest disagreement gives  $M^{\mu}_{J}/M^{s}_{J} \simeq 1.21$. Notice that the last case is associated with  larger values of temperature  ($T=8.87~ {\rm K}$) along with  larger  value in the number density,  $n= 4.99 \times 10^{2} {\rm cm}^{-3}$.

Our analysis was focused on the collapse of overdense region with mass near the Jeans mass  which are associated with an interstellar medium (forming molecular clouds or similar compact objects) and how the viscosity effect can reduce the agglomeration of such matter  within the  extended thermodynamic framework \cite{Mu}. Any way,  as a general rule  we can say that a large collapsing cloud  has the tendency to fragment. Gas above the critical mass ($M>M^{s}_{J}$) will go into a free-fall when there is no significant counter pressure. This is likely, at least initially, since molecular clouds are optically thin so that radiation can easily escape, cooling the cloud and releasing any pressure. The overdensities will therefore contract at different rates. As the molecular cloud collapses, it will break into smaller and smaller pieces in a hierarchical manner, until the fragments reach their Jeans mass. As the density increases, each of these fragments will become
increasingly more opaque and are thus less efficient at radiating away the gravitational potential energy. This raises the temperature of the cloud and inhibits further fragmentation. The fragments might  merge with one another if the increasing angular momentum did not prevent such aggregation.  Then,  compact objects with smaller masses can be formed in the fragmentation process provided other important effects such  the ionization feedback, the radiation pressure, and stellar wind must be taken into consideration along with their numerical analysis \cite{pad}.  Interestingly enough, there are several numerical simulations which show that a massive molecular cloud can be fragmented into several pieces with masses lower than the original Jean mass  leading to the creation of stars and other objects \cite{frag}. Some authors pointed out that the fragmentation process may stop when the mass of the subunit is nearly $M_{\odot}$ \cite{pad}. However, this criterion is not absolut
 e and there is a debate about the possibility of having   compact objects with very low-mass. For instance,  it was reported with the help of the microlensing technique the detection of  objects in  the globular cluster M22  with very low-masses (a fraction of the solar mass), namely  $M=0.13 ^{~+0.03}_{~-0.02}M_{\odot}$ \cite{low}.


\subsection{Expanding Universe}
So far we have handled the case of static universe and analyzed the classical Jeans condition for different astrophysical configurations (mostly stellar clouds) when dissipative effects (such as conduction and viscosity) are included. However, the aforesaid analysis cannot be applicable  to cosmology provided the expansion of the universe was not considered; such key element is essential  for examining the evolution of density inhomogeneities. Then,  let us consider a homogeneous and isotropic spatially flat Universe  described by the Friedmann-Lama\^itre-Robertson-Walker metric $ds^2=(dt)^2-a(t)^2\left(dx^2+dy^2+dz^2\right)$, where $a(t)$ denotes the cosmic scale factor. From Einstein's field equation it follows the so-called  Friedmann and acceleration equations, which  for a pressureless fluid ($p=0$) read
\ben\lb{8a}
\left(\frac{\dot a}a\right)^2=\frac{8\pi G}3\rho,\qquad \frac{\ddot a}a=-\frac{4\pi G}3\rho.
\een
The solution of these equations for the mass density $\rho$ and cosmic scale factor $a$ read:
\ben\lb{8b}
\rho=\rho_{i0}\left(\frac{a_{i0}}{a}\right)^3,~a=a_{i0}\left(6\pi G\rho_{i0}\,t^2\right)^\frac13,~\rho=\frac1{6\pi G t^2}.~~
\een
We will now see what happens to the density perturbation with time. To do so, we write equations of motion for the cosmological fluid. We will consider only one fluid and  volumes of the Universe that are small
compared with the Hubble distance, $|r|\ll H^{-1}$ (sub-horizon scales in a linear regime), so that the Hubble velocities are small, $v \ll c$.
For  the above background, we take that the unperturbed fields are given by
\ben\rho_{0}=\rho_{i0}\left(\frac{a_{i0}}{a}\right)^3, ~~~~ T_{0}=T_{i0}\left(\frac{a_{i0}}{a}\right)^2, 
\\\lb{10}
  v^{i}_0=\frac{\dot a}{a} x_i,~~~~~
\phi_{0}=\frac{2\pi}3G\rho_{0} r^2.~
\een
These equations satisfy (\ref{1a}) and (\ref{1b}) with the constitutive equations (\ref{2a}) identically without the necessity of invoking Jeans swindle approximation. Note that here the subscripts $0$ denote the zero-th order (i.e., homogeneous) solution, not the necessarily values today. On the other hand, the subscripts $i0$ stand for some reference values that we can choose upon our needs. In dealing with the perturbations some cautions must be taken. For instance,
the operator $\nabla_{r}$ in $r$ coordinates must be replaced somehow by a $\nabla_{x}$ operator in $x$ coordinates. To do so, we use that  the comoving coordinates and the physical one are related by the law $x^{i}=r^{i}/a(t)$, implying that  $\nabla_{r}=a^{-1}\nabla_{x}$. Besides,  the physical time derivative  involves the comoving time derivative plus a contribution that accounts for Hubble flow: $\partial_{t}|_{r=cte}=\partial_{t}|_{x=cte}- (v^{i}_0/a)\partial_{i}$. The physical reason  for pointing out the later fact  is the  slight subtlety of the non-commutation of $\partial_{t}|_{r=cte}$ and $\partial_{r}$, which it will be essential  in arriving at the perturbed field equations. In summary,  the time and space derivatives defined from $t$ and $r$ were independent in a static spacetime, however,  this is not the case anymore for  an expanding spacetime.

At this point, we introduce small perturbations by writing $v^{i} = v^{i}_{0} + \delta v^{i}$, $\rho = \rho_{0} + \delta\rho$, $p = p_{0} + \delta p$, and
$\phi = \phi_{0} + \delta\phi,$ where the perturbations are assumed to be small compared with the zero-th order quantities. Furthermore, it is customary  to define the symbol $\delta_{X}$ as a contrast density of a physical quantity, that is,   $\delta X/X_{0}$. The perturbed continuity equation (\ref{1a}) and perturbed Poisson equation (\ref{1b}) can be recast as
\ben\lb{c1}
\dot{\delta}_{\rho} +  \nabla_{i}u^{i}=0,
\een
\ben\lb{p1}
\nabla^{2} \delta \phi= 4\pi G \rho_{0}a^{2}{\delta}_{\rho},
\een
where  ${\delta}_{\rho}\equiv \delta \rho/\rho_{0}$. We have introduced  a new perturbed velocity  $u^{i}= \delta v^{i}/a$, used the relation $\partial_{i} v^{j}_{0}=H\delta^{j}_{i}$ along with the fact that
at  zeroth order in fluctuations (i.e. dropping all perturbations), the usual continuity equation and Poisson equation for unperturbed fields are recovered. Notice that the Laplacian operator refers to derivatives with respect to the comoving coordinates from now on.

Carrying on,  we need to calculate the perturbed balance equation (\ref{1a}) and due to the presence of the pressure tensor term, it seems useful to obtain first its perturbed counter-part.  The perturbed pressure tensor  then is given by
\ben\lb{ppt}
p_{ij}= (p_{0}+ \delta p)\delta_{ij} -\mu_{v}\Big(\partial_{j}u_{i} + \partial_{i}u_{j}-\frac{2}{3}\delta_{ij}\partial_{k}u_{k}\Big).
\een
Leaving aside the viscous contribution in (\ref{ppt}), we can express the perturbed pressure term associated with the isotropic contribution by using the equation of state of an ideal gas. We immediately see that pressure perturbations come with two contributions: $\delta p=p_{0}( \delta_{\rho}+ \delta_{T})$. We recognize the adiabatic term  provided it is proportional to density perturbation, as result it reads $\delta p^{\rm adiab}\equiv c^{2}_{s}\delta \rho$, where $c^{2}_{s}=(3/5)v^{2}_{\rm thermal}$. Nevertheless, there is an additional  non-adiabatic term given by  $\delta p^{\rm non-adiab}\equiv (\delta p/\delta T) \delta T= (k\rho_{0}/m)\delta_{T}$.  As we said before, we are working on sub-horizon scales in a linear regime within the context of Newtonian cosmology which means that we only have one gravitational potential. But,  the use of the pressure tensor with the viscosity term (\ref{ppt}) within the context of a full relativistic perturbation theory would imply
 that there are two different gravitational potentials \cite{DM2}, namely $\Psi$ and $\Phi$.

Now, we are in position to compare our approach with previous articles reported in the literature \cite{velten}, \cite{aqua}. First, we are considering not only non-adiabatic contribution in the perturbed pressure but also the explicit contribution of  $\mu_{v}$-viscosity term (\ref{ppt}). On the other hand, some authors introduced viscosity for exploring the behavior of viscous dark matter in neo-Newtonian cosmology  in a completely different way. They added a bulk viscous (non-adiabatic) term by hand, namely $p \rightarrow p + \Pi_{\rm v}$ \cite{velten}. The lack of a microscopy theory for selecting $\Pi_{\rm v}$ allowed them to choose  a useful parametrization of the bulk viscosity in terms of the density. Of course, such approach led to non-adiabatic perturbation terms quite differently to the one presented in this work. But most importantly, we  preferred to rely on a truncated hierarchy of the full Boltzmann equation provided is a well accepted approach within the context o
 f mechanical statistics (for further detail see Appendix A).  Probably, our line of work is similar to the one employed by Acquaviva \textit{et al} because  they worked within the framework of Newtonian cosmology and they considered the presence of dissipative effects by including a bulk viscosity term which corrects the standard pressure term of Euler equation. Moreover, they added a transport equation for $\Pi_{\rm v}$ based on the Israel and Steward causal theory \cite{aqua}. This is equivalent to promote $\Pi_{\rm v}$  to the rank of another field of the theory which must be determined dynamically.

Coming back to (\ref{ppt}),  the perturbed balance equation becomes
\ben\lb{Eu}
\dot{u}_{i}+ 2H u_{i} + \frac{1}{a^{2}}\Big(\partial_{i}\delta\phi +\frac{\partial_{i}\delta p}{\rho_{0}} -\frac{\mu_{v}}{\rho_{0}}\delta{\cal W}_{i}\Big)=0,
\een
where the extra term is
\ben\lb{Ex}
\delta{\cal W}_{i}=\partial_{jj}u_{i}+ \partial_{ji}u_{j}-\frac{2}{3}\partial_{ik}u_{k}.
\een

We  derive the general evolution equation for the perturbation in temperature. Our starting point is to consider $\epsilon = \epsilon_{0} + \delta\epsilon$, being the  thermal energy density $\epsilon= (3k/2m)T$, so  it is possible to express  $\delta\epsilon= (3k/2m)\delta T$. The evolution equation for $\delta \epsilon$ or $\delta T$ has the gradient term in the heat flux and the contraction of the pressure tensor with the $\partial_{j}v^{i}$ term. For the gradient term, the perturbations are easy to compute, it yields $\rho^{-1}\nabla_{i}q^{i}=-\lambda_{c} a^{-2}(T_{0}/\rho_{0}) \partial_{ii} \delta_{T}$. For the latter case,  we  need to combine (\ref{ppt}) along with the relation  $\partial_{j}v^{i}=a^{-1}(\partial_{j}v_{0i}+ \partial_{j}\delta v_{i})$. Putting all these facts together  leads us
\ben\lb{EdT}
\dot{\delta}_{T}-\frac{5}{2}\frac{\mu(a)}{\rho_{0}a^{2}} \partial_{ii} \delta_{T} + {2\over3}\partial_{i}u_{i}=0.
\een
To obtain (\ref{EdT}), we used  that $\dot{T}_{0}=-2HT_{0}$, $\dot{\delta T}\equiv \dot{T}_{0}\delta_{T}+ T_{0}\dot{\delta}_{T}$, and the relation (\ref{cvs}).

Combining  the time derivative of  (\ref{c1}) with the divergence of (\ref{Eu}), we find
\ben\lb{Mast}
\ddot{\delta}_{\rho}+ 2H\dot{\delta}_{\rho} -\frac{4\mu_{v}}{3a^{2}\rho_{0}}\nabla^{2}\dot{\delta}_{\rho} = \frac{\nabla^{2}}{a^{2}}\Big(\delta \phi+ \delta p\Big).
\een
Expanding the perturbations in Fourier modes as $\delta_{Y} = \sum_{\q}\delta_{Y_{(\q)}}e^{i\q x}$ the Laplacian transforms as $\nabla^{2} \rightarrow -\q^2$, which in turn implies  that the master equation (\ref{Mast}) can be written as
\ben\no
\ddot{\delta}_{\rho (\q)}+\Big(2H+\frac{4\mu_{v} \q^{2}}{3a^{2}\rho_{0}}\Big)\dot{\delta}_{\rho (\q)}
\\\lb{MastFF}
+ \Big(\frac{\q^{2}p_{0}}{a^{2}\rho_{0}}-4\pi G \rho_{0}\Big){\delta}_{\rho (\q)}= S_{\q}.
\een
Here the source term corresponds to  temperature perturbation,  $S_{\q}= (\q^{2}p_{0}/a^{2}\rho_{0}) \delta_{T(\q)}$. Eq. (\ref{MastFF}) tells us that dissipative effects enters into different ways.  The viscosity  alters the usual friction term proportional to $2H$ and the non-adiabaticity pressure term appears as a source for the master equation above. We must emphasize that the physical Jeans scale (length or wave-length) is obtained by demanding that the term  $\propto {\delta}_{\rho (\q)}$ vanishes. Such scale seems to be sensitive to the  equation of state of the gas and consequently it will depend on the thermal velocity of the gas  particles. 

Before presenting  the results of our numerical simulations, we would like to summarize all the master equations expanded in Fourier modes. Using Eqs. (\ref{2a}), (\ref{c1}), (\ref{p1}) and (\ref{Eu}),  we can arrange the whole five-field  system of equations as
\ben\lb{12a}
\dot{\delta}_{\rho (\q)}=- i\q u_{(\q)},
\\\lb{12b}
\dot{\delta}_{T(\q)}+ \frac{5}{2}\frac{\mu_{v}(a)\q^2}{\rho_{0}a^{2}} {\delta}_{T(\q)}=\frac{2}{3}\dot{\delta}_{\rho(\q)},
\\\lb{12bb}
\delta \phi_{(\q)}=- \frac{4\pi G \rho_{0} a^2}{\q^2}\delta_{\rho (\q)},
\\\no\dot{u}_{(\q)}+2H{u}_{(\q)}+ \frac{4\q^2\mu_{v}}{3\rho_{0}a^{2}}{u}_{(\q)} +  \frac{i\q}{a^{2}}\delta \phi_{(\q)}\\\lb{12aa}
 + \frac{i\q}{a^{2}}\frac{p_{0}}{\rho_{0}}[\delta_{T (\q)}+ \delta_{\rho(\q)}]=0.~~~
\een
For practical purposes, we must distinguish between two kinds of (physical) perturbed fields. On the one hand, we have dynamical perturbed fields denoted by $\delta_{\rho (\q)}$ and  ${\delta}_{T(\q)}$ and other derived variables which are associated with some constraint equations, in our model they  are  represented by $u_{(\q)}$ and $\delta \phi_{(\q)}$. The reason for performing such observation is that $\delta \phi_{(\q)}$ can be obtained by simply knowing $\delta_{\rho (\q)}$ through Eq. (\ref{12bb}). The same goes for $u_{(\q)}$ provided it can be obtained from (\ref{12a}) by calculating the time derivative of  $\delta_{\rho (\q)}$. Taking into account the remaining equations, we obtained the following system of equations for the determination of $\delta_{\rho (\q)}$ and $\delta_{T(\q)}$:
\ben\lb{13a}
\ddot\delta_{\rho (\q)}+\left[2\frac{\dot a}a+\frac{4\mu_{v}}{3\rho_{0}}\frac{\q^2}{a^2}\right]\dot\delta_{\rho(\q)} +\omega^{2}_{\q}\delta_{\rho(\q)}=S_{(\q)},
\\\lb{13b}
\dot\delta_{T(\q)}+\frac{5\mu_{v}}{2\rho_{0}}\frac{\q^2}{a^2}\delta_{T(\q)}=\frac23\dot\delta_{\rho(\q)}.~~~
\een
The squared frequency is recast  as   $\omega^{2}_{\q}=[kT_{0}\q^2/ma^{2}-4\pi G\rho_{0}]$ and the source term yields $S_{(\q)}=-(kT_{0}\q^2/ma^2)\delta_{T(\q)}$. The next step is to select  a consistent law for the shear viscosity. To do so, we have to recall that  according to the kinetic theory of gases  the  shear viscosity coefficient for an ideal gas is a function only of the temperature, i.e., $\mu_{v}\propto T^\alpha$, where the exponent of the temperature may range from  $1/2\leq\alpha\leq1$; the value 1/2 is for hard spheres and 1 for soft spheres (Maxwellian molecules) (see e.g. \cite{GK}). Hence, from (\ref{10}) we may write the shear viscosity coefficient in terms of the cosmic scale factor as
\ben\lb{18}
\frac{\mu_{v}}{\mu_{i0}}=\left(\frac{T_{0}}{T_{i0}}\right)^\alpha=\left(\frac{a_{i0}}{a}\right)^{2\alpha}.
\een

We introduce  the adiabatic sound speed $v_s$ along with  the wavelengths $\lambda_0$ and $\lambda_J$ which are  defined by the following expressions:
\ben\lb{16}
v_s=\sqrt\frac{5kT_0}{3m},\qquad\lambda_0=\frac{2\pi a_0}{\q},\qquad\lambda_J=\frac{2\pi v_s}{\sqrt{4\pi G\rho_0}},
\een
Armed with the above definitions, the system of equations (\ref{13a}) and (\ref{13b}) can be rewritten as
\ben\no\tau^2\delta_{\rho(\q)}''+\frac43\left[1+\sqrt{\frac23}\mu_*\left(\frac{\lambda_J}{\lambda_0}\right)^2
\tau^{\frac{5-4\alpha}{3}}\right]\tau\delta'_{\rho(\q)}
\\\lb{19a}
-\frac23\left[\delta_{\rho(\q)}-\frac35\left(\frac{\lambda_J}{\lambda_0}\right)^2
\frac{\delta_{\rho(\q)}+\delta_{T(\q)}}{\tau^\frac23}\right]=0,~~~
\\\lb{19b}
\tau\left(\delta'_{T(\q)}-\frac23\delta'_{\rho(\q)}\right)
+\frac5{\sqrt6}\mu_*\left(\frac{\lambda_J}{\lambda_0}\right)^2\tau^{\frac{5-4\alpha}{3}}\delta_{T(\q)}=0.
\een
Above we have used  explicitly that the universe is dominated by matter so the relationships $\dot a/a=2/3t$ and $\rho_{0}=1/(6\pi Gt^2)$ hold. Further, we  introduced the  dimensionless time $\tau=t\sqrt{6\pi G\rho_{i0}}$ and the prime  stands for derivative  with respect to $\tau$.  We remark that the above system of equations  is not easy to tackle for several reasons. We start by recalling that they form a coupled system of equations and it is even worse than it looks provided  the Bessel equation  (\ref{19a}) is non-homogeneous. Nevertheless, we are going to obtain the leading contributions due to the viscosity effects  by using a series of physical approximations. Further,  we  will confirm the aforesaid solution by solving  the coupled system  numerically.

Let us search for a solution of the system of equations (\ref{19a}) and (\ref{19b}). For simplicity, we consider the case where $\alpha=1/2$, i.e., the case of hard-spheres. First, we introduce an auxiliary contrast density $\delta_{\beta (\q)}$ in order to decouple  the solution of this system of equations. In doing so,  we define \ben\lb{a2}
\delta_{\beta(\q)}=\delta_{T(\q)}-\frac23\delta_{\rho(\q)},
\een
so that (\ref{19b}) can be rewritten as
\ben\lb{a3}
\delta_{\beta(\q)}'+\frac5{\sqrt6}\mu_*\left(\frac{\lambda_J}{\lambda_0}\right)^2\left(\frac23\delta_{\rho(\q)}
+\delta_{\beta(\q)}\right)=0.
\een
The solution of (\ref{a3}) can be splitted as the sum of two contributions as follows:
\ben\no\delta_{\beta(\q)}=e^{-\frac5{\sqrt6}\mu_*\left(\frac{\lambda_J}{\lambda_0}\right)^2\tau} -\frac23e^{\frac5{\sqrt6}\mu_*\left(\frac{\lambda_J}{\lambda_0}\right)^2\tau} \\\lb{a4}
 \times\int^\tau dt \delta_{\rho(\q)}(t)\frac5{\sqrt6}\mu_*\left(\frac{\lambda_J}{\lambda_0}\right)^2 e^{-\frac5{\sqrt6}\mu_*\left(\frac{\lambda_J}{\lambda_0}\right)^2t}.~~
\een
From now on, we shall investigate solutions that depart from the case without dissipation ($\mu_*=0$) by considering the coefficient $\epsilon=\frac5{\sqrt6}\mu_*\xi^{2}_{J}$ as a small quantity. Note that in this case the shear viscosity coefficient for  small wavelengths $\xi_{J}>1$ should be smaller than that for  large wavelengths $\xi_{J}<1$. Hence we can approximate the solution (\ref{a4}) as
\ben\lb{a5}
\delta_{\beta (\q)}=1-\epsilon\tau- \frac23\epsilon\int^\tau \delta_{\rho (\q)}(t') dt'.
\een
Given the physical nature of $\mu_*$, we can consider the density contrast as perturbative series  in the small (control) parameter $\epsilon$, namely $\delta_{\rho (\q)}=\delta_{1\rho(\q)}+\epsilon\delta_{2\rho(\q)}$. Replacing the latter ansatz in  (\ref{19a}) we obtain two differential equations,  one for $\delta_{1\rho (\q)}$ and another for $\delta_{2\rho (\q)}$:
\ben\no
\tau^2\delta_{1\rho(\q)}''+\frac43\tau\delta_{1\rho(\q)}'-\frac23\left(1-\xi^{2}_{J}\tau^{-\frac23}\right)\delta_{1\rho(\q)}
\\\lb{a7a}
=-\frac25\xi^{2}_{J}\tau^{-\frac23},
\\\no\tau^2\delta_{2\rho (\q)}''+\frac43\tau\delta_{2\rho(\q)}'-\frac23\left[1-\xi^{2}_{J}\tau^{-\frac23}\right]\delta_{2\rho(\q)}
\\\no
-\frac25\xi^{2}_{J}\tau^{-\frac23}\left(\tau+\frac23\int^\tau \delta_{1\rho(\q)}(t') dt'\right)
\\\lb{a7b}
+\frac8{15}\tau^2\delta_{1\rho(\q)}'
=0.
\een
The homogeneous solution of (\ref{a7a}) is given in terms of Bessel functions of first kind $J_{\pm\frac52}\left(\frac{\sqrt6\xi_J}{\tau^\frac13}\right)$ and its general solution reads
\ben\no
\delta_{1\rho(\q)}=t^{-\frac16}\left[C_1 J_{+\frac52}\left(\frac{\sqrt6 \xi_{J}}{\tau^\frac13}\right)+C_2 J_{-\frac52}\left(\frac{\sqrt6 \xi_{J}}{\tau^\frac13}\right)\right]
\\\lb{a8a}
-\frac35\left[1+\xi_{J}\tau^\frac23\right].
\een
We now calculate the  leading terms of (\ref{a8a}) under  the assumption that $\xi_{J}<1$, so we are looking at the regime where the Jeans instability has already begun. In that situation, (\ref{a8a}) exhibits to contributions, namely $\delta_{1\rho(\q)} \simeq \delta_{1g}\tau^{2/3}+ \delta_{1d}\tau^{-1}$ being $\delta_{1g}$ and $\delta_{1d}$ two constants.  In order to simplify the instability analysis, we have restricted our attention to the growing  mode only, thus it is valid to consider $\delta_{1\rho (\q)} \simeq \delta_{1g}\tau^{2/3}$. Replacing the latter proposal in (\ref{a7b}) and neglecting the term  proportional to $\xi^{2}_{J}$,  it is possible to  put Eq. (\ref{a7b})  in a more suitable shape for extracting the growing and decaying modes. Then,  Eq. (\ref{a7b})  becomes
\ben\lb{a7bb}
\tau^2\delta_{2\rho (\q)}''+\frac43\tau\delta_{2\rho(\q)}'-\frac23\delta_{2\rho(\q)}+ \frac{16}{45}\delta_{1g} \tau^{\frac53}=0.~~
\een
 whose solution  is given by
\ben\lb{a7xx}
\delta_{2\rho(\q)}=-\frac{2}{15}\delta_{1g} \tau^{\frac53} + \delta_{2g}\tau^{\frac23}+ \delta_{2d}\tau^{-1},
\een
where $\delta_{2g}$ and $\delta_{2d}$ are two constants. In other words, the viscosity term  generates an additional growing mode which is proportional to  $\tau^{\frac53}$. Taking into account that the total density contrast is $\delta_{\rho(\q)}=\delta_{1\rho}+\epsilon\delta_{2\rho}$, we can conclude that the total density contrast has two growing modes: the standard one given by $\tau^{\frac23}$ plus another that appears as a correction in the limit of small viscosity, $\tau^{\frac53}$. Of course, the latter term is sub-leading in the limit of very small viscosity and the standard expression for the density contrast is recovered for a universe dominated by dust-like matter. Having obtained the leading behavior in the contrast density, we are able to determine the leading behavior of some derived quantities. From (\ref{12a}), we observe that $u_{(\q)} \propto \delta'_{\rho (\q)}$ so the peculiar velocity exhibits also a principal growing mode ($\tau^{2/3}$), a subleading gr
 owing mode ($\propto \epsilon \tau^{5/3}$) and a decaying mode ($ \propto \tau^{-1}$). Using (\ref{12bb}), we arrive at the conclusion that the perturbed gravitational field $\delta \phi_{(\q)} \propto  a^{-1}\delta_{\rho (\q)}$, indicating that the modes include three terms, one is proportional to a constant, another goes like $\tau^{-5/3}$ and the subleading correction scales as $\epsilon \tau^{2/3}$. The latter fact shows us the gravitational potential remains constant in the limit of vanishing viscosity,  as one could expect.
\begin{figure}[h!]
\includegraphics[width=16cm]{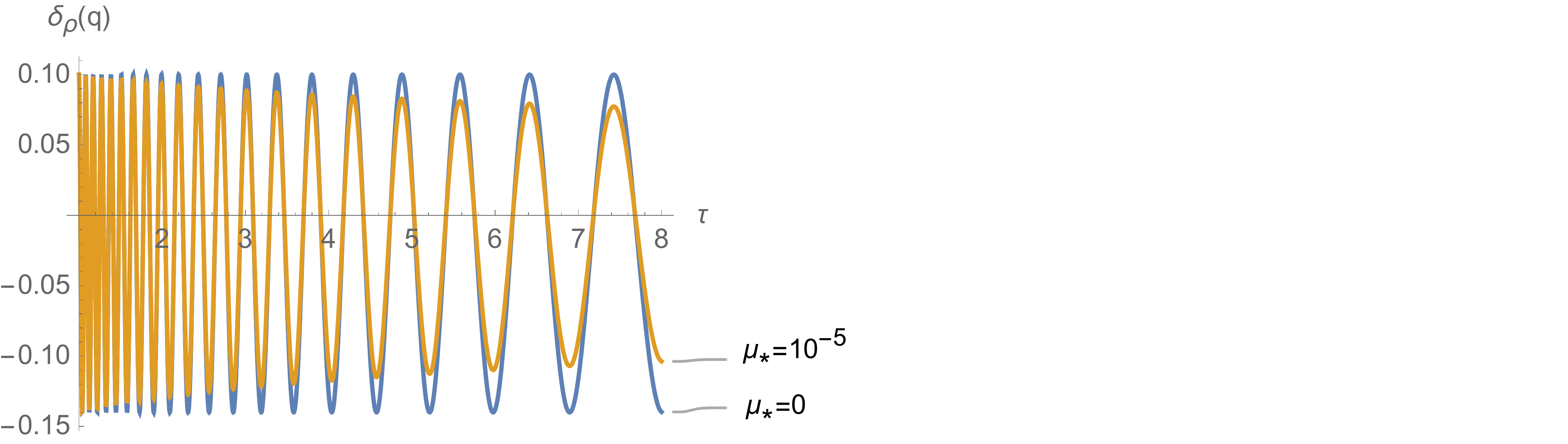}
\caption{ Time evolution of the density contrast for small wavelengths with  $\alpha=1$ and $\xi_{J}=100$.}
\lb{fig2}
\end{figure}
\begin{figure}[h]
\includegraphics[width=12cm]{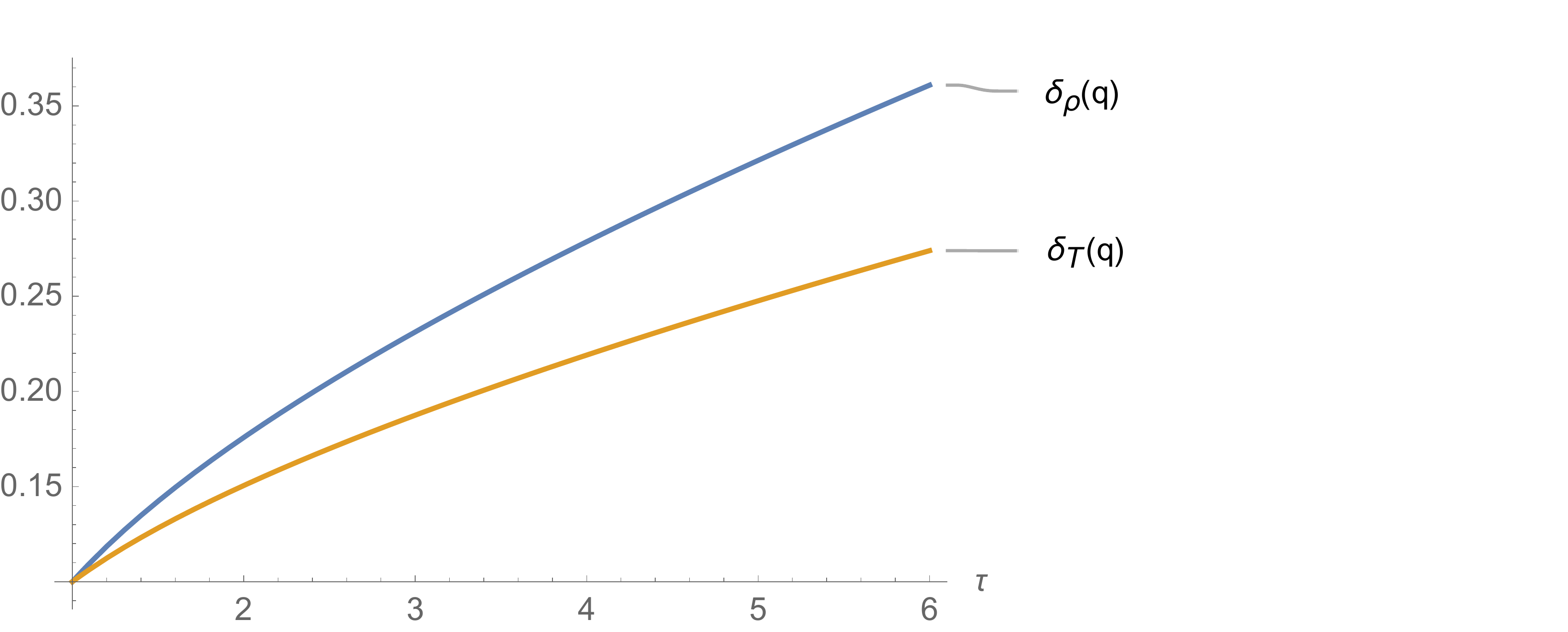}
\caption{Time evolution of the density contrast and temperature contrast for large wavelengths with $\alpha=1$ and $\xi_{J}=1/2$.}
\lb{fig3}
\end{figure}

As a simple way to confirm our previous analytical results we  are going to solve the system of equation (\ref{19a}) and (\ref{19b}) without any approximation. Then, we will numerically determine the conditions for having Jeans instability when the universe  is  dominated by a pressureless fluid and  the viscosity is nonzero. To do so,  the coupled system of  differential equations (\ref{19a}) and (\ref{19b}) for the density and temperature contrasts were solved numerically  for the cases of small ($\xi_{J}>1$) and large ($\xi_{J}<1$) wavelengths. Without loss of generality,  we considered the following initial conditions $\delta_{\rho(\q})(1)=\delta'_{\rho(\q)}(1)=\delta_{T(\q)}(1)=10^{-1}$.  In  Figs. \ref{fig2} and \ref{fig3} the time evolution of the density and temperature contrasts are plotted, respectively. In both cases  we  chosen $\alpha=1$ -- which refers to soft particle model -- but  for other values of $1/2\leq\alpha\leq1$ the behavior of the curves  seems to be
   quite similar, so we are not loosing generality.  The density and the temperature contrasts shown in Fig. \ref{fig2} for the case of small wavelengths $\xi_{J}=100$ have the same behavior and represent harmonic waves. When the shear viscosity coefficient vanishes the density and temperature contrasts evolve with time without damping, while when the shear viscosity is present the harmonic waves  are damped and by increasing the value of this coefficient the damping becomes more accentuated. Fig. \ref{fig3} shows the solutions for large wavelengths $\xi_{J}=0.5$ where the density and temperature contrasts grow with time and correspond to Jeans instability. We infer from this figure that when the shear viscosity is present the evolution with time of the density and temperature contrasts are less accentuated than the one without viscosity. This can be understood by noting that in the presence of viscous and thermal effects the
  dissipation in the energy makes the process of structure formation less effective. Regarding the physical impact of  the shear viscosity coefficient, we have noticed that it has a strong  influence on the temperature contrast while  the density contrast seems to be less sensitive to it.

\section{Jeans instability from a thirteen-field theory}
In this section, we will examine an enlarged model where the pressure tensor and heat flux vector are endowed with their own master equations. The physical motivation for extending our previous approach can be explained as follows. The balance equations (\ref{1a}) and (\ref{1b}) with the Navier-Stokes and Fourier (\ref{2a}) constitutive equations lead to a parabolic system of equations which imply infinite speeds for the  heat propagation. This was known in the literature as the heat paradox (see \cite{Mu}). It can be  solved by introducing a hyperbolic system of equations and the most simple system of equations is to consider the thirteen fields of mass density $\rho$, velocity $v_i$, pressure tensor $p_{ij}$ and heat flux vector $q_i$ (cf. Appendix A). For an ideal gas these balance equations can be obtained from the Grad method of kinetic theory (see e.g. \cite{GK}) or from the so-called extended thermodynamic theory (see \cite{Mu}). The field equations for the mass densit
 y $\rho$, velocity $v_i$, pressure tensor $p_{ij}$ and heat flux vector $q_i$ read

\ben\lb{e1a}
\frac{\partial \rho}{\partial t}+v_i\frac{\partial \rho}{\partial x_i}+\rho\frac{\partial  v_i}{\partial x_i}=0,
\\\no
\frac{\partial v_i}{\partial t}+v_j\frac{\partial v_i}{\partial x_j}+\frac{kT}{m\rho}\frac{\partial \rho}{\partial x_i}+\frac{k}{m}\frac{\partial T}{\partial x_i}   
\\\lb{e1ai}
+\frac1\rho\frac{\partial \sigma_{ij}}{\partial x_j}+\frac{\partial \phi}{\partial x_i}=0,
\\\no
\frac{3k}{2m}\left(\frac{\partial T}{\partial t}+v_i\frac{\partial T}{\partial x_i}\right)+\frac1\rho\frac{\partial q_i}{\partial x_i}+\frac{k}{m}T\frac{\partial v_i}{\partial x_i}   \\\lb{e2}
+\frac{\sigma_{ij}}\rho\frac{\partial v_i}{\partial x_j}=0,
\\\no
\frac{\partial \sigma_{ij}}{\partial t}+v_k\frac{\partial \sigma_{ij}}{\partial x_k} +\sigma_{ij}\frac{\partial  v_k}{\partial x_k}+\frac25\left(\frac{\partial q_i}{\partial x_j}+\frac{\partial q_j}{\partial x_i}\right)
\\\no
-\frac{4}{15}\frac{\partial q_r}{\partial x_r}\delta_{ij} +\sigma_{ik}\frac{\partial  v_j}{\partial x_k}+\sigma_{jk}\frac{\partial  v_i}{\partial x_k} -\frac23\sigma_{kr}\frac{\partial  v_k}{\partial x_r}\delta_{ij} 
\\\lb{e2i}
+\rho\frac{k}{m}T\left(\frac{\partial  v_i}{\partial x_j}+\frac{\partial  v_j}{\partial x_i}-\frac23\frac{\partial  v_r}{\partial x_r}\delta_{ij}\right)=-\nu\sigma_{ij},
\\\no
\frac{\partial q_i}{\partial t}+v_j\frac{\partial q_i}{\partial x_j} +\frac75q_i\frac{\partial  v_k}{\partial x_k}+\frac25q_j\frac{\partial  v_j}{\partial x_i}+\frac75q_j\frac{\partial  v_i}{\partial x_j}   
\\\no
+\left(\frac{p}{\rho}\delta_{ij}-\frac{\sigma_{ij}}\rho\right)\frac{\partial  \sigma_{jk}}{\partial x_k}+\frac{p}\rho\sigma_{ij}\left(\frac{5}{2T}\frac{\partial T}{\partial x_k}-\frac1\rho\frac{\partial \rho}{\partial x_k}\right)  
\\\lb{e4}
+\frac{5k}{2m}p\frac{\partial T}{\partial x_i} =-\frac23\nu q_i.
\een
This  system of equations together with Poisson equation (\ref{1b}) is a closed system for the determination of  $\rho$,  $v_i$,  $p_{ij}$ and  $q_i$. In order to obtain  (\ref{e2}) and (\ref{e4}) we have decomposed the pressure tensor into a pressure isotropic contribution $p$ plus a stress tensor $\sigma_{ij}$, i.e., $p_{ij}=p\delta_{ij}+\sigma_{ij}$. We have written down two evolution equations, the first one refers to its trace and the second one to $\sigma_{ij}$. Equation (\ref{e4}) refers to the evolution equation for the heat flux vector. The master equations for $\sigma_{ij}$ and $q_i$ involve  constitutive terms which were taken as functions of the thirteen basic fields according to Grad's method (see Appendix A). The constitutive terms refer to a third-order moment $p_{ijk}$, a contracted fourth-order moment $q_{ij}=p_{ikkj}$ and two production terms $P_{ij}$ and $P_i$. According to kinetic theory of gases their expressions in a linearized theory read (see e.g. \cite{GK})
\ben
p_{ijk}=\frac25\left(q_i\delta_{jk}+q_j\delta_{ik}+q_k\delta_{ij}\right),~~ P_{ij}=-\nu \sigma_{ij},\\\lb{yy}
q_{ij}=\frac{5p^2}{2\rho}\delta_{ij}+\frac{7p}{2\rho}\sigma_{ij},\qquad P_i=-\frac23\nu q_i.
\een
Here $\nu$ is a collision frequency which, according to the kinetic theory of gases, is related with the shear viscosity coefficient and pressure by $\nu=p/\mu_{v}$.

\subsection{Static Universe}

As  in the previous section we search for solutions of the system (\ref{1b}) and (\ref{e1a}) -- (\ref{e4}) in the form of plane waves. At equilibrium the density, temperature and gravitational potential assume constant values while the velocity, stress and heat flux vanish.  The perturbations out of equilibrium refer to plane waves  of small amplitudes with frequency $\omega$ and wavenumber $\k$ propagating in the $x$ direction.

\ben\lb{e6}
\rho=\rho_0+ \delta\rho e^{i\left(\k x-\omega t\right)},\quad v_x=\delta v_xe^{i\left(\k x-\omega t\right)},~~
\\
T=T_0+\delta Te^{i\left(\k x-\omega t\right)},\quad  \delta \sigma_{xx}=\delta\sigma e^{i\left(\k x-\omega t\right)},
\\\lb{e7}
\sigma_{ij}= \delta \sigma_{xx}{\rm diag}  [1, -\frac{1}{2}, -\frac{1}{2}], \quad q_{x}=\delta qe^{i\left(\k x-\omega t\right)}~~
\\\lb{e8}
\phi=\phi_0+ \delta\phi e^{i\left(\k x-\omega t\right)}.
\een
Above we have also that the constant values of density, temperature and gravitational potential satisfy  (\ref{e1a}) -- (\ref{e4}) but not the Poisson equation (\ref{1b}), so that we have to impose Jeans swindle which says that the
(\ref{1b}) is valid only for the perturbations.

From the insertion of the representations (\ref{e6}) -- (\ref{e8}) into the (\ref{e1a}) -- (\ref{e4}) and  (\ref{1b}) we get a system of algebraic equations for the amplitudes and in order to have a solution for this system the determinant of the coefficients of the amplitudes must vanish. This implies the dispersion relation:
\ben\no\omega_*^5+\frac{i}\mu_*\omega^4+\left(1-\frac{78}{25}\k_*^2-\frac6{25\mu_*^2}\right)\omega_*^3
\\\no
+\frac{i}{\mu_*}\left(1-\frac{48}{25}\k_*^2\right)\omega_*^2 -\frac6{25}\left[\frac{1-\k_*^2}{\mu^2_*}\right.
\\\lb{e10a}
\left.+\frac{11\k_*^2}2\left(1-\frac{9\k_*^2}{11}\right)\right]\omega_*-\frac{3i\k_*^2}{25\mu_*}\left(5-3\k_*^2\right)=0.~~~~
\een
Here again by considering a vanishing shear viscosity $\mu_*=0$ we get Jeans solution (\ref{7a}).

If we consider  big wavelengths $\lambda_J/\lambda<1$ it follows  from (\ref{e10a}) five values for the dimensionless frequency
\ben\lb{e10b}
\omega_{*1}=-i\left[\frac{5\mu_*}2\left(\frac{\lambda_J}{\lambda}\right)^2+\dots\right],~~~~~~~~~~~~~~~
\\\lb{e10c}
\omega_{*2}=-i\left[1-\frac{3(1-3\mu_*)}{2(3-5\mu_*)}\left(\frac{\lambda_J}{\lambda}\right)^2-\dots\right],~~~~~~~~~~~~~
\\\lb{e10d}
\omega_{*3}=-i\left[\frac{2}{5\mu_*}-\frac{9\mu_*}{10}\left(\frac{\lambda_J}{\lambda}\right)^2-\dots\right],~~~~~~~~~~~~~~~~
\\\lb{e10d2}
\omega_{*4}=-i\left[\frac{3}{5\mu_*}-\frac{4\mu_*(33-50\mu_*^2)}{5(9-25\mu_*^2)}\left(\frac{\lambda_J}{\lambda}\right)^2-\dots\right],
\\\lb{e10e}
\omega_{*5}=i\left[1-\frac{3(1+3\mu_*)}{2(3+5\mu_*)}\left(\frac{\lambda_J}{\lambda}\right)^2-\dots\right].~~~~~~~~~~~
\een
All dimensionless frequencies are pure imaginary with $\omega_{*1}$ to $\omega_{*4}$ implying  decaying perturbations in time whereas $\omega_{*5}$ grows with time and is identify with the Jeans instability. All frequencies show up  a dependence on dissipative phenomena through the dimensionless shear viscosity coefficient $\mu_*$.

For  small wavelengths $\lambda_J/\lambda>1$ it follows from (\ref{e10a}) one or two sets of complex conjugate solutions for the dimensionless frequency which correspond to harmonic waves in time and three or one solutions which refer to the decay of perturbations   in time, respectively.


As part of our numerical analysis, we investigate the behavior of  the real part of $\exp(-i\omega t)$  as a function of time $t$ (\ref{e10a}) for the ratio  $\xi_{J}=2$ (small wavelengths) for different values of the dimensionless shear viscosity coefficient. The solutions are harmonic waves in time and the one without the shear viscosity ($\mu_*=0$)  represents a non-damped wave which corresponds to the Jeans solution as was pointed out previously. The other two solutions with $\mu_*\neq0 $ represent damped harmonic waves and we infer from this figure that the increase of the shear viscosity coefficient increase the wave damping. For large values of the shear viscosity coefficient  a strong damping occurs so that the harmonic character of the wave is lost. It was found that the solutions of (\ref{e10a}) for the ratio  $\xi_{J}=1/2$ (large wavelengths) represents the growth of the perturbations with time, showing up  the Jeans instability. One can infer from such analysis th
 at the Jeans solution ($\mu_*=0$) grows more rapid with respect to time than those two where the shear viscosity coefficients do not vanish. This behavior can be explained on the basis of an energy loss due to the dissipative effects so that the growth of the perturbations are damped. Note that the growth with time of the perturbation by increasing the shear viscosity coefficient becomes less accentuated. A similar behavior as that described above can also be obtained from the five-field theory of subsection IIA, but here we have used field equations of hyperbolic type rather than of parabolic type so that the heat paradox is avoided.

\subsection{Expanding Universe}

As in subsection IIC we analyze the solution of the thirteen field equations (\ref{e1a}) -- (\ref{e4}) coupled with Poisson equation (\ref{1b}) in form of plane waves of small amplitudes in an expanding Universe. The background values for the fields of mass density $\rho_{0}$, temperature $T_{0}$, velocity $ v_i$ and gravitational potential $\phi_{0}$ are given by (\ref{10}), whereas those values for the fields of stress and heat flux  are considered to be zero, i.e., $ \sigma^{0}_{ij}=q^{0}_i=0$. The background values for the fields are superposed by longitudinal plane waves of time dependent small amplitudes, namely $Y=Y_{0}(t)+ \sum_{\q}\delta Y_{(\q)} e^{i\q x}$. In our case, $Y_{0}(t)\equiv \{\rho_{0}, v_{x0}, T_{0}, \phi_{0} \}$ provided $q_{x0}=0= \sigma^{0}_{xx}=0$, while the perturbed fields are given by $\delta Y_{(\q)}=\{\delta\rho(t), \delta v(t), \delta T(t), \delta q(t), \delta \sigma (t) \}$, where we  used that  the shear tensor has some symmetries, namely onl
 y the diagonal components are non-zeros  and the trace of such tensor vanishes, $ 2\sigma_{yy}=2\sigma_{zz}=-\sigma_{xx}$, as a result we only need to introduce just one function called $\delta \sigma(t)$. Replacing the latter proposals into the balance equations (\ref{e1a}) --  (\ref{e4}) and (\ref{1b}) and linearization it follows a coupled system of differential equations for the amplitudes:
\ben\lb{e12ai}
\dot{\delta}_{\rho(q)}=-i\q u_{(\q)},
\\\no
\dot{u}_{(\q)}+2H u_{(q)}+ \frac{i\q}{a^{2}}\Big(\frac{kT_{0}}{m} (\delta_{\rho (\q)}+ \delta_{T (\q)}) + \delta \phi_{(\q)}\Big)
\\\lb{e12a}
=-\frac{i\q}{a^{2}\rho_{0}}\delta\sigma_{(\q)},~~~
\\\lb{e12aii}
\frac{3}{2}\dot{\delta}_{T(\q)}+\frac{i\q}{a\rho_{0}} \frac{m}{kT_{0}}\delta q_{(\q)}+ i\q u_{(\q)} =0,
\\\no
\dot{\delta\sigma}_{(\q)} + 5H\delta\sigma_{(\q)} + \frac{i\q8}{15a}\delta q_{(\q)}+\nu\delta\sigma_{(\q)}
\\\lb{e12b}
+\frac{i\q p_{0}}{3}u_{(\q)}=0,
\\\lb{e12b1}
\delta \phi_{(\q)}=- \frac{4\pi G \rho_{0} a^2}{\q^2}\delta_{\rho (\q)},
\\\no
\dot{\delta q}_{(\q)} + 6H \delta q_{(\q)}+ \frac{i\q kT_{0}}{ma}\delta \sigma_{(\q)}+ \frac{i\q 5kT_{0}}{2ma}\delta_{T(\q)}
\\\lb{e12c}
+\frac{2\nu}{3}\delta q_{(\q)}=0.
\een
Here also the background solution satisfies equations (\ref{e1a}) --  (\ref{e4}) and (\ref{1b}) identically without the use of Jeans swindle. The above coupled system can be written in a simpler form by introducing the dimensionless time $\tau=t\sqrt{6\pi G \rho_{i0}}$  and the density, temperature, stress and heat contrasts defined as $\delta_{\rho(\q)}= \delta \rho_{(\q)}/\rho_{0}$, $\delta_{T(\q)}= \delta T_{(\q)}/T_{0}$, $\delta_{\sigma(\q)}= \delta \sigma_{(\q)}/\sigma_{0}$ and  $\delta_{q (\q)}= \delta q_{(\q)}/q_{0}$. The functions used to arrange dimensionless heat flux and shear tensor are given by
\ben\lb{e13}
\sigma_{0}=\frac{kT_{0}\rho_{0}}{m}, ~~~~q_{0}=\frac{\alpha_{0} kT_{0}\rho_{0}a}{i\q m}\een
Furthermore, we recall that the collision frequency is related with the pressure and the coefficient of shear viscosity by $\nu=p/\mu_{v}$ while the latter is given in terms of the cosmic scale factor by (\ref{18}) and $a=a_0(6\pi G\rho_{i0}t^2)^\frac13$. Hence we have that
\ben\lb{e14}
\nu=\frac{p_{0}}{\mu_{v}}=\frac35\frac{\sqrt{4\pi G\rho_0}}{\mu_*\tau^{\frac23(5-2\alpha)}}
\een
where $\alpha_{0}=6\pi G\rho_{i0}$.
From equation (\ref{e12a}) we obtain that $\dot\delta_{\rho (\q)}=-i\q u_{(\q)}$ and we can rewrite the coupled system of equations (\ref{e12a}) -- (\ref{e12c}) in terms of the dimensionless time $\tau$ as
\ben\no
\tau^2\delta''_{\rho(\q)}+\frac43\tau\delta'_{\rho(\q)}-\frac23\delta_{\rho(\q)}   
\\\lb{e15a}
+\frac2{5\tau^\frac23}
\xi^2_{J}\left(\delta_{\rho(\q)}+\delta_{T(\q)}+\delta_{\sigma(\q)}\right)=0,
\\\lb{e15b}
\tau\left(\delta'_{T(\q)}-\frac23\delta'_{\rho(\q)}\right)+\frac49\delta_{\sigma(\q)}
+\frac23\tau\delta_{q(\q)}=0,
\\\no
\tau\delta'_{\sigma(\q)}-\frac43\delta_{\sigma(\q)}+\frac8{15}\tau\delta_{q(\q)}
-\frac43\tau\delta'_{\rho(\q)}   \\\lb{e15c}
=-\frac{3}{5\mu_*}\sqrt{\frac23}\tau^{-(7-4\alpha)/3}\delta_{\sigma(\q)},
\\\no
\tau\delta'_{q(\q)}+\frac43\delta_{q(\q)}-\frac25\xi^2_{J}
\tau^{-\frac53}\delta_{\sigma (\q)}-\xi^2_{J}\delta_{T(\q)}
\\\lb{e15d}
=-\frac{3}{5\mu_*}\sqrt{\frac23}\tau^{-(7-4\alpha)/3}\delta_{q(\q)},~~
\een
where as was previously introduced the primes refer to a derivation with respect to $\tau$. Equations (\ref{e15a}) -- (\ref{e15d}) represent a coupled system of differential equations for the determination of the mass density $\delta_\rho$, temperature $\delta_T$, stress $\delta_\sigma$ and heat flux $\delta_q$ contrasts.

The coupled system of differential equations (\ref{e15a}) -- (\ref{e15d})  was solved numerically with the following initial conditions $\delta_{\rho(\q)}(1)=\delta'_{\rho(\q)}(1)=\delta_{T(\q)}(1)=\delta_{\sigma(\q)}(1)=\delta_{q(\q})(1)=10^{-1}$ and for the case of Maxwellian particle ($\alpha=1$). In the upper panel of Fig.  \ref{fig4} are plotted the time evolution of the density, temperature, stress  and heat flux contrasts, which behave as damped harmonic waves for small wavelengths. For large wavelengths the same contrasts are shown as function of the time in the lower panel of this figure. It is interesting to note that the density and temperature contrasts grow with time, while the stress and heat flux contrasts  heavily decay with time. The growth of the density contrast is associated with Jeans instability.

\begin{figure}[h]
\includegraphics[width=12cm]{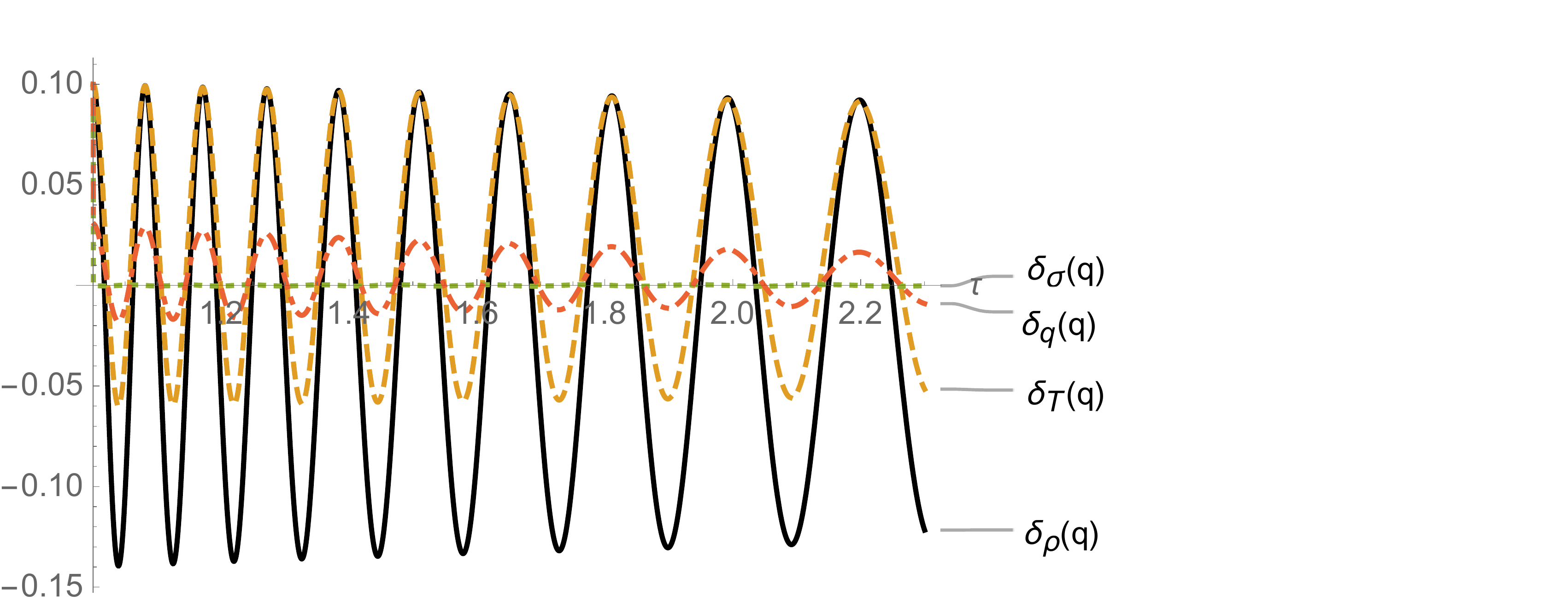}\hskip1cm\includegraphics[width=12cm]{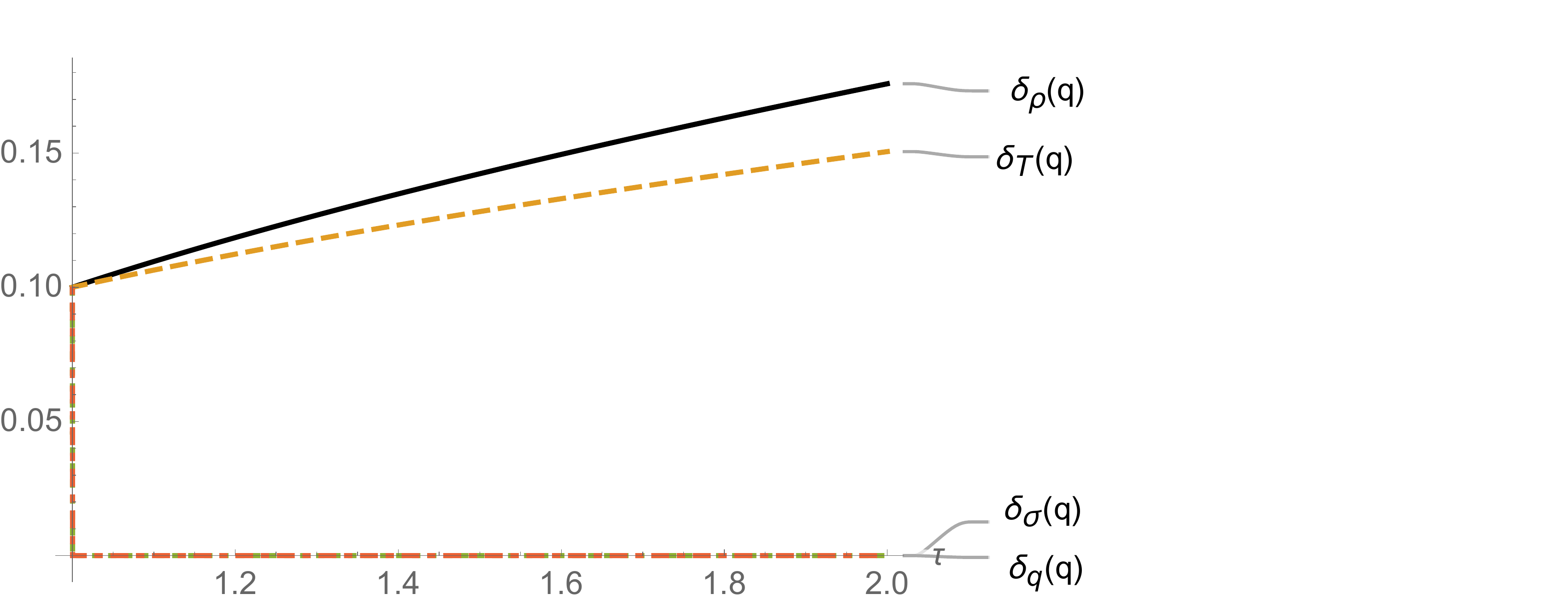}
\caption{Upper level: Time evolution of the density (straight line), temperature (dashed line), stress (dotted line) and heat flux (dot-dashed line) as damped  harmonic waves  for small wavelengths $\xi_{J}=100$ and dimensionless shear viscosity $\mu_*=10^{-5}$. Lower level: Time evolution of the density (straight line), temperature (dashed line), stress (dotted line) and heat flux (dot-dashed line)   for large wavelengths $\xi_{J}=0.5$ and dimensionless shear viscosity $\mu_*=10^{-5}$.}
\lb{fig4}
\end{figure}

\section{Summary}
We have explored the gravitational amplification of inhomogeneities  within the context of five and thirteen (thermodynamic) field theories for static and expanding universes. We provided two simple physical pictures of how the thermal effects along with viscous ones  do affect the growth of instabilities. In this scheme, we  have started by calculating the dispersion relation in both models, obtaining that the changes introduced by viscosity and the heat flux  considerably modified the propagating modes.  In the first part, we only considered static universes so the Jeans instability  only involved two elements: pressure and gravity. We then studied  Jeans instability mechanism for a universe dominated by pressureless matter, generalizing earlier results on the gravitational instability in the presence of heat flux and viscosity. Besides, we calculated the Jeans mass for stellar configurations (mostly molecular clouds) and their behaviors with the number density, temperature
 , and viscosity. The outcome of such analysis confirmed that astrophysical configurations with viscosity need greater Jeans mass to collapse.

We devoted our effects to extract the main physical insight about the coupled system of contrast density and contrast temperature by focusing  in the model of hard sphere and Maxwellian particles. In the small wavelengths regime, increasing the amount of shear viscosity into the system forces the harmonic perturbations to damp faster
whereas in the opposite limit larger values of shear viscosity lead to smaller values of density and
temperature contrasts.  In the limit of  large wavelength ($\lambda_{J}<\lambda_{0}$), our analytical results indicate that viscous effects are subleading ones provided they generate an additional growing modes proportional to $\tau^{5/3}$ which becomes important for intermediate values of viscosity parameter, whereas in the limit of very small viscosity they can be neglected and the standard behavior of contrast density is recovered ($\propto\tau^{2/3}$) for a matter dominated universe.  

In order to further understand the Jeans instability with dissipative effects, we have
directed our attention to the thermodynamic  thirteen-field model. The aforesaid scheme represents a hyperbolic system of differential equation characterized by  two related parameters, the shear viscosity and the collision frequency. We found that
dispersion relation becomes a polynomial in the frequency with two order higher in relation to
the five-field theory, indicating that the effects associated with the shear viscosity and heat flux become
non-trivial. Regarding the dynamical evolution of the density, temperature, stress and heat flux
contrasts for a universe dominated by pressureless matter, we obtain also damped harmonic waves
for small wavelengths. In the case of large wavelengths, the density and temperature contrasts grow
with time (due to the Jeans mechanism) while the stress and heat flux contrasts decay with time.

Even though, we were able to make some progress in the analysis of Jeans instability with dissipative effects, showing that thermal and viscous effects contribute to  reduce the ability of the system to aggregate matter in a cooperative way and therefore the structure formation may be less effective.  It would be interesting to see whether analytic/numerical results for the growth of perturbations can be obtained by inspecting  the full Boltzmann equation for  one baryon-like specie plus  self-interacting dark matter. In this way,  one could address the issue of warm dark matter coupled to baryons and by doing so one could take into account thermal along with viscous effects  as a consequence of the microscopic dynamic of the system.  We leave the investigation of this question for future work.

\appendix
\section{Balance equations and constitutive equations within the framework of  the Boltzmann equation}
In this work we have analyzed the Jeans instability of a viscous and heat conducting ideal non-relativistic gas through the use of the balance equations of ordinary and extended thermodynamics. Although the balance equations can be determined from a phenomenological theory, the constitutive equations adopted in this work were taken from a kinetic theory based on the Boltzmann equation. In the following we shall sketch the procedure used in the kinetic theory in order to obtain the balance and constitutive equations. For more details the reader should consult a book on the Boltzmann equation (see e.g. \cite{GK}).

The starting point is the Boltzmann equation which is the space-time evolution of the one-particle distribution function $f({\bf x},{\bf \xi},t)$ in the phase space spanned by the space $\bf x$ and velocity $\bf \xi$ coordinates of the gas particles. In the presence of a gravitational field $\phi$ the Boltzmann equation reads \cite{GK}.
\ben\lb{a1}
\frac{\partial f}{\partial t}+\xi_i\frac{\partial f}{\partial x_i}-\frac{\partial \phi}{\partial x_i}\frac{\partial f}{\partial \xi_i} =\mathcal{Q}(f,f).
\een
The right-hand side of the above equation takes into account the binary elastic collisions between the particles and $\mathcal{Q}(f,f)$ denotes the collision operator of the Boltzmann equation. If $(\xi,\xi_1)$ and  $(\xi^\prime,\xi_1^\prime)$ denote the pre  and post collisional velocities of two particles at a binary collision the  collision operator reads \cite{GK}.
\ben\no
\mathcal{Q}(f,f)&=&\int\big[f({\bf x},{\bf \xi_1^\prime},t)f({\bf x},{\bf \xi^\prime},t)
\\
&-&f({\bf x},{\bf \xi_1},t)f({\bf x},{\bf \xi},t)\big]g b db d\epsilon d{\bf \xi}_1.
\een
At this point, it is illustrative to show that the above expression of the collision operator already contained the scattering cross section associated with the collisional process amongst classical particles. To do so,  one can notice   that for simple scattering (elastic), an ``event'' is defined as the deflection of one particle into a range $d\Omega$ of solid angles about some observation direction $\bar{\Omega}$. Using polar coordinates, one can write  the differential solid angle as  $d\Omega= \sin \chi d\chi d\epsilon$, where 
 $\epsilon$ denotes an azimuthal  angle and $\chi$  is the scattering angle which defines the scattering process.   Notice that, in general, for a particular interaction potential $V (r)$ between the particles, the scattering angle $\chi$ depends on relative velocity $g=\vert \xi_1-\xi\vert$  and “impact parameter $b$ (miss distance). If there is a one-to-one relation between $b$ and $\chi$, the particles that will be deflected in the referred range of angles have impact parameters between $b(\chi)$ and $b(\chi) + db(\chi)$. By simple geometry,  we can write $\sigma d\Omega= b(\chi) db(\chi) d\epsilon$, so the differential cross section of the scattering process is given by 
\ben\no
\sigma(\chi, g)= \frac{b(\chi)}{\sin \chi}\Big|\frac{db(\chi)}{d\chi}\Big|.
\een 
The latter equation tells us that the collision operator encodes the information of the scattering process in the term  $g  b(\chi) d b(\chi) d\epsilon=g \sigma(\chi, g) d\Omega$. Further, the physical information about the interacting potential amongst the particles is hidden in the generic expression of the scattering angle $\chi$ (see \cite{GK}). We would like to   remark that  we will be dealing  with classical particles which are treated as hard spheres provided is useful for obtaining an explicit expressions to describe classical gases. It is a tractable model and also describes in good detail collisions at high energies, where the attractive wells are negligible compared with the kinetic energy. In this model, particles of radius $r_0=d/2$ interact with a central potential that is infinitely hard when particles meet at a distance $d$ . When the interactions are described by the hard sphere  of diameter $d$, one can simplify the scattering section provided  the impact parameter and the deflection angle are related by the next expression: $b = d \cos (\chi/2)$;  then the scattering section is given by $\sigma_{\rm total}=d^2/4$ \cite{GK}. As expected, total scattering cross section is proportional to the projected area of the sphere. Let us end this remark on the scattering section by mentioning why the classical approach is possible.  Since collisions occur at atomic distance, their rigorous analysis requires Quantum Mechanics. Specifically, this is so whenever the distance of closest approach the scattering section (of the order of $\sqrt{\sigma_{\rm total}}$ ) is
comparable to or less than the Broglie wavelength for the relative momentum $\hbar/p$. Putting $p \simeq \sqrt{\mu_{\rm red}kT}$ with $\mu_{\rm red}=m_{1}m_{2}/(m_{1}+m_{2})$, quantum effects dominate when $\sigma_{\rm total} < {\hbar}^{2}/(\mu_{\rm red}kT)$.  As long as  the latter condition holds,  classical dynamics can be used in the calculation of the scattering section as we have done before,  and then the overall collisional effects  which appear in the collision operator of the Boltzmann equation are obtained on firm grounds. For instance,  n-n collisions, $\mu_{\rm red}>m_{\rm H}=1.7 \times 10^{-27}{\rm kg}$, and $T=3000 {\rm K}$ this requires $\sigma_{\rm total}<10^{-22} {\rm m}^{2}$.

Now, we show how the macroscopic fields in kinetic theory are given as mean values with respect to the one-particle distribution function. The mass density $\rho$, velocity $\bf v$, specific internal energy $\varepsilon$, pressure tensor $p_{ij}$ and heat flux vector $q_i$ are defined by 
\ben\lb{f1}
\rho=\int m fd^3\xi,\qquad v_i=\frac1\rho\int m\xi_i fd^3\xi,
\\\lb{f2}
\varepsilon=\frac1{2\rho}\int m\xi^2 fd^3\xi,
\\\lb{f3}
p_{ij}=\int m\xi_i\xi_j fd^3\xi,\qquad q_i=\int \frac{m}2\xi^2\xi_i fd^3\xi,
\een
where $m$ is the particle mass.

For the case of ordinary thermodynamics the balance equations  are obtained by multiplying the Boltzmann equation (\ref{a1}) with the mass $m$, momentum $m\xi_i$ and energy $m\xi^2/2$ of a particle and integration over all values of the velocity $d^3\xi$. Following this methodology we arrive at the balance equations of mass (\ref{1a})$_1$ and of velocity (\ref{1a})$_2$. For the determination of the balance equation for the specific internal energy one has to subtract from the balance equation for the total energy the corresponding balance equation for the kinetic energy of the particles $\rho v^2/2$. Note that the collision operator does not furnish any contribution to these balance equations, since mass, momentum and energy are conserved in an elastic collision.

The Maxwellian distribution function
\ben
f_0=\frac{m^\frac12\rho}{(2\pi kT)^\frac32}\exp\left[\frac{m({\bf \xi-v})^2}{2kT}\right],
\een
refers to the equilibrium solution of the Boltzmann equation (\ref{a1}). It is a function of the mass density $\rho$, velocity $\bf v$ and temperature $T$ of the gas. Above $k$ is the Boltzmann constant. From the knowledge of the equilibrium distribution function it is easy to determine the pressure and the specific internal energy in terms of the mass density and temperature, namely, $p=\rho k T/m$ and $\varepsilon=3kT/2m$.

The non-equilibrium distribution function is obtained from the Boltzmann equation through the Chapman-Enskog method (see e.g. \cite{GK}). In this method the distribution function is written as a Maxwellian distribution function $f_0$ plus a small deviation, namely $f=f_0(1+\varphi)$. In order to determine the  deviation $\varphi$, the Maxwellian distribution is inserted into the left-hand side of the Boltzmann equation and the time derivatives of the mass density, velocity and temperature are eliminated by the use of the Euler equations. The expression for the the distribution function  $f=f_0(1+\varphi)$ is inserted on the right-hand side of the Boltzmann equation and the products of the deviations are neglected. Hence  it is found that the deviation  is a function of the velocity and temperature gradients, namely 
\ben\no
f&=&f_0\bigg\{1-\frac{3\mu_{v}}{2p}\bigg[\left(\frac{m\xi^2}{2kT}-\frac52\right)\frac{\xi_i}T\frac{\partial T}{\partial x_i}
\\\lb{ce}
&+&\frac{m}{3kT}\xi_i\xi_j\left(\frac{\partial v_i}{\partial x_j}+\frac{\partial v_j}{\partial x_i}-\frac23\frac{\partial v_k}{\partial x_k}\delta_{ij}\right)\bigg]\bigg\}.
\een
Here  the shear viscosity coefficient  is
\ben
\mu_{v}=\frac5{16}\frac1{\Omega^{(2,2)}}\sqrt{\frac{mkT}\pi},
\een
which is given in terms of the collision integral 
\ben
\Omega^{(2,2)}=\int_0^\infty\int_0^\infty e^{-\gamma^2}\gamma^7(1-\cos^2\chi)bdbd\gamma.
\een
Above $\gamma=\sqrt{m/2kT}g$ and $\chi$ is the scattering angle. For a potential of hard spheres $b=d\cos(\chi/2)$ where $d$ is the diameter of the particle, so that $\Omega^{(2,2)}=d^2$ and the shear viscosity coefficient becomes
$\mu_{v}=\frac5{16}\frac1{d^2}\sqrt{\frac{mkT}\pi}.$

 If we substitute the non-equilibrium distribution function (\ref{ce}) into the definitions of the pressure tensor and heat flux vector (\ref{f3}) and integrate over all particle velocities we get the Navier-Stokes and Fourier constitutive equations (\ref{2a}), where the relation between the thermal conductivity and shear viscosity coefficients is given by $\lambda_{c}=15k\mu_{v}/4m$ for all spherically symmetric interaction potentials. Furthermore, the shear viscosity coefficient does not depend on the mass density, but only on the temperature. This dependence can be written as $\mu_{v}\propto T^\alpha$ where the value of $\alpha$ depends on the interaction potential, for example $\alpha=1/2 $ for hard sphere potential and $\alpha=1$ for soft sphere potential (Maxwellian particles).

In the case of extended thermodynamics we base on  Grad's 13-moment method and apart from the balance equations of the mass density and velocity,  we introduce the balance equations for the pressure tensor and for the heat flux vector. These balance equations are obtained from the multiplication of the Boltzmann equation by $m\xi_i$ and $m\xi^2\xi_i/2$ and integration over all values of the velocity $d^3\xi$, respectively. They read
\be
\frac{\partial p_{ij}}{\partial t}+\frac{\partial(p_{ijk}+p_{ij}v_k)}{\partial x_k}+p_{ki}\frac{\partial v_j}{\partial x_k}+p_{kj}\frac{\partial v_i}{\partial x_k}=P_{ij},\;
\ee{aa2}
\ben\no
\frac{\partial q_i}{\partial t}+\frac{\partial(q_{ij}+q_iv_j)}{\partial x_j}+p_{ijk}\frac{\partial v_j}{\partial x_k}+q_j\frac{\partial v_i}{\partial x_j}
\\\lb{aa3}
-\frac{p_{ki}}\rho\frac{\partial p_{kj}}{\partial x_j}-\frac12\frac{p_{rr}}\rho\frac{\partial p_{ij}}{\partial x_j}=Q_i.
\een
In the above equations we have introduced new moments of the distribution function, namely
\ben\lb{a5}
p_{ijk}=\int m\xi_i\xi_j\xi_k f d^3\xi,\qquad q_{ij}=\int \frac{m}2 \xi^2\xi_i f d^3 \xi,
\een
and the production terms
\be\lb{a6}
P_{ij}=\int m\xi_i\xi_j \mathcal{Q}(f,f)d^3\xi, \; Q_i=\int \frac{m}2\xi^2\xi_i\mathcal{Q}(f,f)d^3\xi.
\ee{zz}

In order to close the system of equations we have to express the new moments of the distribution function and production terms as functions of the 13-moments $\rho, v_i, T, \sigma_{ij},q_i$, where $\sigma_{ij}$ is the symmetric and traceless part of the pressure tensor $p_{ij}$. This is attained by using Grad's distribution function which is given in terms of these quantities, namely (see e.g. \cite{GK})
\be
f=f_0\left\{1+\frac{\rho}{2p^2}\left[\sigma_{ij}\xi_i\xi_j+\frac45q_i\xi_i\left(\frac{m\xi^2}{2kT}-\frac52\right)\right]\right\}.
\ee{gr}
 Insertion of the Grad's distribution function (\ref{gr}) into the definition of the new moments (\ref{a5})  and production terms (\ref{a6}) and integrating over all the velocities $d^3\xi$ we arrive at the linearized constitutive equations (\ref{yy}).
Once these constitutive equations are known, the insertion of them into the balance equations (\ref{aa2}) and (\ref{aa3}) implies eqs. (\ref{e2i}) and (\ref{e4}), once we decompose the pressure tensor $p_{ij}$ in (\ref{aa2}) in its trace $p_{rr}=3p$ and its symmetric traceless part $\sigma_{ij}$.

\acknowledgments
We would like to thank the referee for making useful suggestions, which helped improve the article.
G. M. K and M. G. R are supported by Conselho Nacional de Desenvolvimento Cient\'ifico e Tecnol\'ogico (CNPq)- Brazil.
F.T is supported by Coordena\c{c}\~ao de Aperfei\c{c}oamento de Pessoal de N\'ivel Superior (CAPES)-Brazil.

\end{document}